\documentstyle[11pt,psfig,aaspp4]{article}
\begin{document}

\lefthead{Psaltis et al.}
\righthead{Horizontal Branch Oscillations}

\title{On the Magnetospheric Beat-Frequency and
Lense-Thirring Interpretations of the Horizontal
Branch Oscillation in the Z Sources}

\author{Dimitrios Psaltis\altaffilmark{1}, 
Rudy Wijnands\altaffilmark{2},
Jeroen Homan\altaffilmark{2},
Peter G.\ Jonker\altaffilmark{2},\\
Michiel van der Klis\altaffilmark{2,3},
M.\ Coleman Miller\altaffilmark{4},
Frederick K.\ Lamb\altaffilmark{5},
Erik Kuulkers\altaffilmark{6,7},\\
Jan van Paradijs\altaffilmark{2,8},
and Walter H.\,G.\ Lewin\altaffilmark{9}}

\altaffiltext{1}{Harvard-Smithsonian Center for
Astrophysics, 60 Garden St., Cambridge, MA 02138, 
U.S.A.; dpsaltis@cfa.harvard.edu}

\altaffiltext{2}{Astronomical Institute ``Anton
Pannekoek'', University of Amsterdam and Center for
High Energy Astrophysics, Kruislaan 403, NL-1098 SJ
Amsterdam, The Netherlands; rudy, homan, peterj,
michiel, jvp@astro.uva.nl}

\altaffiltext{3}{Department of Astronomy, University 
of California at Berkeley, Berkeley, CA 94720}

\altaffiltext{4}{Department of Astronomy and
Astrophysics, University of Chicago, 5640 South Ellis
Avenue, Chicago, IL 60637, U.S.A.; 
miller@bayes.uchicago.edu}

\altaffiltext{5}{Departments of Physics and Astronomy, 
University of Illinois at Urbana-Champaign, 1110 W. Green 
St., Urbana, IL 61801, U.S.A.; f-lamb@uiuc.edu}

\altaffiltext{6}{Astrophysics, University of Oxford,
Nuclear and Astrophysics Laboratory, Keble Road,
Oxford, OX1 3RH, United Kingdom}

\altaffiltext{7}{Present address: Space Research
Organization Netherlands, Sorbonnelaan 2, 3584 CA
Utrecht, \&\ Astronomical Institute, Utrecht
University, P.O.~Box 80000, 3507 TA, Utrecht, The
Netherlands; E.Kuulkers@sron.nl}

\altaffiltext{8}{Department of Physics, University of
Alabama at Huntsville, Huntsville, AL 35899}

\altaffiltext{9}{Department of Physics and Center for
Space Research, Massachusetts Institute of
Technology, Cambridge, MA 02139; lewin@space.mit.edu}

\begin{center}
 To appear in the {\em Astrophysical Journal\/}.
 \end{center}

\newpage

\begin{abstract}
 Three types of quasi-periodic oscillations (QPOs) have
been discovered so far in the persistent emission of
the most luminous neutron-star low-mass X-ray
binaries, the Z sources:  $\sim$10--60~Hz horizontal
and $\sim$6--20~Hz normal/flaring branch oscillations,
and $\sim$200--1200~Hz kilohertz QPOs, which usually
occur in a pair. Here we study the horizontal branch
oscillations and the two simultaneous kilohertz QPOs,
which were discovered using the {\it Rossi X-ray
Timing Explorer\/}, comparing their properties in five
Z sources with the predictions of the magnetospheric
beat-frequency and Lense-Thirring precession models. 

We find that the variation of the horizontal-branch
oscillation frequency with accretion rate predicted by
the magnetospheric beat-frequency model for a purely
dipolar stellar magnetic field and a
radiation-pressure-dominated inner accretion disk is
consistent with the observed variation. The model
predicts a universal relation between the horizontal
branch oscillation, stellar spin, and upper kilohertz
QPO frequencies that agrees with the data on five Z
sources. The model implies that the neutron stars in
the Z sources are near magnetic spin equilibrium, that
their magnetic field strengths are $\sim
10^9$--$10^{10}$~G, and that the critical fastness
parameter for these sources is $\gtrsim 0.8$. If the
frequency of the upper kilohertz QPO is an orbital
frequency in the accretion disk, the magnetospheric
beat-frequency model requires that a small fraction of
the gas in the disk does not couple strongly to the
stellar magnetic field at 3--4 stellar radii but
instead drifts slowly inward in nearly circular orbits
until it is within a few kilometers of the neutron
star surface.

The Lense-Thirring precession model is consistent with
the observed magnitudes of the horizontal branch
oscillation frequencies only if the moments of inertia
of the neutron stars in the Z sources are $\sim 4$--5
times larger than the largest values predicted by
realistic neutron-star equations of state. If instead
the moments of inertia of neutron stars have the size
expected and their spin frequencies in the Z sources
are approximately equal to the frequency separation of
the kilohertz QPOs, Lense-Thirring precession can
account for the magnitudes of the horizontal branch
oscillation frequencies only if the fundamental
frequency of the horizontal branch oscillation is at
least four times the precession frequency. The trend
of the correlation between the horizontal branch
oscillation frequency and the frequency of the upper
kilohertz QPO is not consistent with this model.

We argue that the change in the slope of the
correlation between the frequency of the horizontal
branch oscillation and the frequency of the upper
kilohertz QPO, when the latter is greater than 850~Hz,
is directly related to the varying frequency
separation of the kilohertz QPOs.
 \end{abstract}



\section{INTRODUCTION}

The discovery in the Z-type low-mass X-ray binaries
(LMXBs) of $\sim$10--60~Hz quasi-periodic oscillations
(QPOs) with centroid frequencies that are positively
correlated with mass accretion rate (van der Klis et
al.\markcite{vdk85}\ 1985; see van der
Klis\markcite{vdK89} 1989 for a review) has led to a
significant improvement in our understanding of such
systems. These horizontal-branch oscillations (HBOs),
which are named after the branch in X-ray color-color
diagrams where they appear, were the first
rapid-variability phenomena discovered in LMXBs and
have played a key role in organizing the complex
phenomenology of these sources that has emerged over
the past decade (see, e.g., Hasinger \& van der
Klis\markcite{HK89} 1989; van der Klis 1989). 

Very soon after the discovery of the HBO, its centroid
frequency was identified with the difference between
the Keplerian frequency at the radius where the
neutron star magnetosphere couples strongly to the gas
in the accretion disk and the spin frequency of the
neutron star (Alpar \& Shaham\markcite{AS85} 1985;
Lamb et al.\markcite{Letal85}\ 1985; Shibazaki \&
Lamb\markcite{SL87} 1987). This magnetospheric
beat-frequency interpretation of the HBO was found to
agree well with the observed properties of the HBOs,
including the dependence of the HBO frequency on X-ray
countrate (Alpar \& Shaham\markcite{AS85} 1985; Lamb
et al.\markcite{Letal85}\ 1985; Ghosh \&
Lamb\markcite{GL92} 1992), the existence of correlated
low-frequency noise (Lamb et al.\ 1985; Shibazaki \&
Lamb 1987), and the absence of any detectable QPO with
a frequency equal to the Keplerian frequency at the
magnetic coupling radius (Lamb 1988). The
magnetospheric beat-frequency model predicted that the
neutron stars in the Z sources have spin frequencies
$\sim 200$--350~Hz and magnetic field strengths $\sim
10^9$--$10^{10}$~G (see Alpar \& Shaham 1985; Ghosh \&
Lamb 1992; Wijnands et al.\markcite{Wetal96}\ 1996),
consistent with the hypothesis that these stars are
the progenitors of the millisecond rotation-powered
pulsars (Alpar \& Shaham 1985; see Alpar et
al.\markcite{Aetal82}\ 1982;  Radhakrishnan \&
Shrinivasan\markcite{RS82} 1982). The neutron star
properties inferred from the magnetospheric
beat-frequency model have subsequently been shown to
be consistent with the magnetic field strengths
inferred from models of the X-ray spectra of the Z
sources (Psaltis, Lamb, \& Miller\markcite{PLM95}
1995; Psaltis \& Lamb\markcite{PL98} 1998) and with
the 290--325~Hz spin frequencies inferred from the
frequency separation of the two simultaneous QPOs with
frequencies $\sim 1$~kHz (hereafter the kilohertz
QPOs) and from the high-frequency oscillations
observed during type~I X-ray bursts in other
neutron-star LMXBs (Strohmayer et
al.\markcite{Setal96} 1996;  Miller, Lamb, \&
Psaltis\markcite{MLP98} 1998;  Miller\markcite{M99}
1999). If the upper kilohertz QPO is an orbital
frequency in the inner disk, the magnetospheric
beat-frequency model of the HBO requires that a small
fraction of the gas in the accretion disk must
penetrate to radii smaller than the radius where it
initially couples to the stellar magnetic field (van
der Klis et al.\ 1997; see Miller et
al.\markcite{MLP98}\ 1998 for a discussion), because
observations show that the kilohertz QPOs are present
at the same time as the HBO (see, e.g., Wijnands \&
van der Klis\markcite{WK98} 1998). In addition to the
HBOs, the magnetospheric beat-frequency model has been
used to explain successfully similar QPOs observed in
accretion-powered pulsars, where the neutron star spin
frequency can be measured directly and the magnetic
field strength can be estimated from the accretion
torque, providing a stringent test of the model (see,
e.g., Angelini, Stella, \& Parmar\markcite{ASP89}
1989; Ghosh\markcite{G96} 1996; Finger, Wilson, \&
Harmon\markcite{FWH96} 1996).

Recently, Stella \& Vietri\markcite{SV98} (1998) have
proposed an alternative HBO mechanism, motivated by
concern about whether orbiting gas can penetrate
inside the magnetic coupling radius in the Z sources.
In this model, the magnetic field of the neutron star
plays no role in generating the HBO. Instead, the HBO
observed in the Z sources and the power-spectral peaks
with frequencies $\sim20$--60~Hz seen in some atoll
sources (see, e.g., Ford \& van der Klis 1998)  are
both generated by nodal (Lense Thirring and classical)
precession of a tilted ring of gas at a special radius
in the inner disk. Stella \& Vietri suggested that the
nodal precession frequency of the ring is visible in
X-rays because of the changes in the Doppler shift of
radiation from blobs orbiting in the ring, changes in
occultations by such blobs, or the changing aspect of
the ring seen by an observer. Subsequently, Markovi\'c
\& Lamb (1998) studied the normal modes of the inner
disk and showed that typically $\sim10$ high-frequency
nodal precession modes are weakly damped. Nodal
precession has also been proposed by Cui, Zhang, \&
Chen\markcite{CZC98} (1998; see also
Ipser\markcite{I96} 1996) as an explanation for the
QPOs observed in black hole candidates. If the HBO is
generated by nodal precession at the same radius in
the accretion disk where orbital motion generates the
kilohertz QPO, as proposed by Stella \& Vietri (1998),
then the HBO frequency, the neutron star spin
frequency, and the frequency of the upper kilohertz
QPO should satisfy a specific relation. The shape of
this relation was shown to be consistent with
observations of the HBO and kilohertz QPO frequencies
observed in the Z sources GX~17$+$2 and GX~5$-$1
(Stella\markcite{S97} 1997; Stella \& Vietri 1997,
1998; Morsink \& Stella\markcite{MS99} 1999), although
the predicted precession frequencies were found to be
smaller than the observed HBO frequencies.

In this paper we use data on five Z sources obtained
using the {\it Rossi X-ray Timing Explorer\/} (RXTE)
to investigate further the origin of the HBO. All of
these data have been fully reported elsewhere. The
sources we consider are GX~17$+$2 (Wijnands et
al.\markcite{Wetal97}\ 1997; Homan et
al.\markcite{Hetal98}\ 1998), GX~5$-$1 (Wijnands et
al.\markcite{Wetal98b}\ 1998b), GX~340$+$0 (Jonker et
al.\markcite{Jetal98}\ 1998), Cyg~X-2 (Wijnands et
al.\markcite{Wetak98a}\ 1998a), and Sco~X-1 (van der
Klis et al.\markcite{Ketal97}\ 1997). In all of these
sources the HBO and two kilohertz QPOs have been
observed simultaneously. In GX~349$+$2, the sixth
originally identified Z source, no HBO has so far been
detected simultaneously with the kilohertz QPOs
(Kuulkers \& van der Klis\markcite{KK98} 1998; Zhang,
Strohmayer, \& Swank\markcite{ZSS98} 1998). Therefore,
we cannot include this source in the present study. We
investigate the magnetospheric beat-frequency model in
\S2 and the Lense-Thirring precession model in \S3,
comparing their predictions with the available data.
In \S4 we summarize our conclusions and their
implications for the properties of the neutron stars
in the Z sources. Finally, we characterize the
correlations between the various frequencies in a
model-independent way in Appendix~A, in order to
facilitate comparison of the present data with future
data or other theoretical models.

\section{THE MAGNETOSPHERIC BEAT-FREQUENCY 
INTERPRETATION}

\subsection{Model Predictions and Comparison with 
Observations}

In the magnetospheric beat-frequency model of the HBO
(Alpar \& Shaham 1985; Lamb et al.\ 1985), the centroid
frequency $\nu_{\rm HBO}$ of the HBO is identified with
the beat between the Keplerian frequency $\nu_{{\rm
K},m}$ at the radius $r_{m}$ where the neutron star
magnetic field couples strongly to the gas in the
accretion disk and the spin frequency $\nu_{s}$ of the
neutron star. The frequency of this beat is
 \begin{equation}
 \nu_{\rm MBF}=\nu_{{\rm K},m}-\nu_{s}\;.
 \label{bfm}
 \end{equation}
 Ghosh \& Lamb (1992) computed the dependence of
$\nu_{\rm MBF}$ on the stellar mass and magnetic moment
and the accretion rate for a variety of simple models
of the inner accretion disk. They found that if the
coupling radius is in an asymptotic region of the
disk, then
 \begin{equation}
 \nu_{\rm MBF} \simeq \nu_{\rm K,0} M^{\gamma}
    \mu_{27}^{\beta}\left(\frac{\xi\dot{M}} 
   {\dot{M}_{\rm E}}\right)^{\alpha} - \nu_{s}\;,
 \label{mag}
 \end{equation}
 where $\mu_{27}$ is the magnetic moment of the
neutron star in units of $10^{27}$~G~cm$^{3}$, $M$ is
its gravitational mass in units of solar masses,
$\dot{M}$ is the mass accretion rate, and
$\dot{M}_{\rm E}$ is the Eddington critical mass
accretion rate onto a neutron star of 10~km radius;
the proportionality constant $\nu_{\rm K,0}$ and the
exponents $\alpha$, $\beta$, and $\gamma$ in
equation~(\ref{mag}) are different for different
models of the inner accretion disk and are listed in
Table~1. The dimensionless parameter $\xi$ describes
the fraction of the mass flux through the inner disk
that couples to the stellar magnetic field at $r_{m}$
and is introduced here to allow for the possibility
that some of the gas in the disk does not couple to
the stellar magnetic field at $r_{m}$ but instead
penetrates to smaller radii, as required if the upper
kilohertz QPO is an orbital frequency in the inner
disk (Miller et al.\markcite{MLP98}\ 1998); $\xi$ may
depend on the mass accretion rate.

In the magnetospheric beat-frequency model of the HBO,
the steep dependence of $\nu_{\rm HBO}$ on the mass
accretion rate inferred from the {\it EXOSAT\/} data
implies that $\nu_{\rm MBF}$ is small compared to
$\nu_{{\rm K},m}$ and hence that the spin frequencies
$\nu_{s}$ of the neutron stars in the Z sources are
very close to, but less than, $\nu_{{\rm K},m}$ (Alpar
\& Shaham 1985; Lamb et al.\ 1995; Ghosh \& Lamb
1992). Stated differently, the magnetospheric
beat-frequency model of the HBO requires that the
neutron stars in the Z sources be near magnetic spin
equilibrium. Indeed, the 200--350~Hz spin frequencies
predicted by the model (see Ghosh \& Lamb 1992) are
much larger than the $\approx 20$--50~Hz HBO
frequencies, as required. The similarity of the
Z-source spin frequencies predicted by the
magnetospheric beat-frequency model to the spin
frequencies inferred from the separation frequencies
of the kilohertz QPOs in the Z sources lends further
support to the model (Miller et al.\markcite{MLP98}\
1998) and to its implication that the neutron stars in
the Z sources are near magnetic spin equilibrium
(White \& Zhang 1997; Psaltis \& Lamb 1998). If they
are, then their spin frequencies are given by (Ghosh
\& Lamb 1979, 1992)
 \begin{equation}
  \nu_{s} \simeq
   \omega_c \nu_{\rm K,0}M^{\gamma}\mu_{27}^{\beta}
    \left[
     \frac{\langle(\xi\dot{M})^{\alpha}\rangle}
          {(\dot{M}_{\rm E})^{\alpha}}
   \right] \;,
  \label{nueqm}
 \end{equation}
 where $\omega_c$ is the critical fastness parameter
and the angle brackets indicate an average over a time
interval equal to the timescale on which the accretion
torque changes the spin.

Combining equations~(\ref{mag}) and~(\ref{nueqm}) and
identifying $\nu_{\rm HBO}$ with $\nu_{\rm MBF}$,
we find
 \begin{equation}
  \frac{\nu_{\rm HBO}}{\nu_{s}}+1 =
   \frac{1}{\omega_c}
   \left[
   \frac{(\xi\dot{M})^\alpha}
       {\langle (\xi\dot{M})^\alpha \rangle}
   \right] \;.
  \label{nuind}
 \end{equation}
 Equation~(\ref{nuind}) shows that the magnetospheric
beat-frequency model of the HBO predicts that
$\nu_{\rm HBO}/\nu_{s} + 1$ should be $\approx
1/\omega_c$ and hence only slightly larger than unity.
 The Z sources are thought to be accreting at
near-Eddington accretion rates (see Lamb\markcite{L89}
1989; Hasinger \& van der Klis\markcite{HK89} 1989).
The inner accretion disk in these sources is therefore
expected to be radiation-pressure-dominated, in which
case $\alpha \approx 0.2$ (see Table~1). Hence, if
$\xi$ depends only weakly on the instantaneous
accretion rate $\dot M$, then $\nu_{\rm HBO}/ \nu_{s}
+ 1$ will also depend only weakly on $\dot M$ and
possibly also on the magnetic field and mass of the
neutron star, through the dependence of $\omega_c$ on
these quantities (any such dependence is expected to
be weak).

According to the magnetospheric beat-frequency model,
the HBO frequency $\nu_{\rm HBO}$ is related to the
frequency $\nu_2$ of the upper kilohertz QPO only
indirectly, through the dependence of both frequencies
on the mass accretion rate. In all sources in which
kilohertz QPOs have so far been discovered, $\nu_2$
increases with inferred mass accretion rate (see,
e.g., van der Klis et al.\ 1996;  Strohmayer et al.\
1996; van der Klis 1998). Here we explore the
consequences of the simple ansatz $\nu_2 = \nu_0
(\dot{M}/\dot{M}_{\rm E})^\lambda$, where $\nu_0$ and
$\lambda$ are constants that are specific to each
source and may depend on the mass and magnetic field
strength of the neutron star. This relation, with
$\lambda \approx 1$, is consistent with kilohertz QPO
observations of several atoll sources, provided that
the observed countrate from an atoll source is
proportional to the mass accretion rate (see, e.g.,
Strohmayer et al.\ 1996; Ford et al.\ 1997). In all
the Z sources, $\nu_2$ is consistent with being
$\simeq1200$~Hz when they are accreting at
near-Eddington rates, which implies $\nu_0 \approx
1200$~Hz, independent of the expected modest
differences in the masses and magnetic field strengths
of these neutron stars. Using this simple ansatz,
equation~(\ref{mag}) can be written
 \begin{equation}
 \nu_{\rm HBO}+\nu_{s}=A_1 \nu_2^{\alpha/\lambda}\;,
 \label{magHBO}
 \end{equation}
 where
 \begin{equation}
  A_1\equiv \xi^\alpha \nu_{\rm K,0} 
    M^{\gamma}\mu_{27}^\beta \nu_0^{-\alpha/\lambda}\;,
 \end{equation}
 and equation~(\ref{nuind}) becomes
 \begin{equation}
  \frac{\nu_{\rm HBO}}{\nu_{s}}+1
 = A_2^{-1} (\nu_2/\nu_0)^{\alpha/\lambda}\,
  \label{nuindHBO}
 \end{equation}
 where 
 \begin{equation}
 A_2 \equiv
  \omega_c
    \left[
     \frac{\langle(\xi\dot{M})^{\alpha}\rangle}
       {(\xi\dot{M}_{\rm E})^{\alpha}}
    \right] \;.
 \end{equation}
 The inferred value of $A_2$ therefore provides an
estimate of the critical fastness $\omega_c$.

\begin{figure}[t]
 \centerline{
\psfig{file=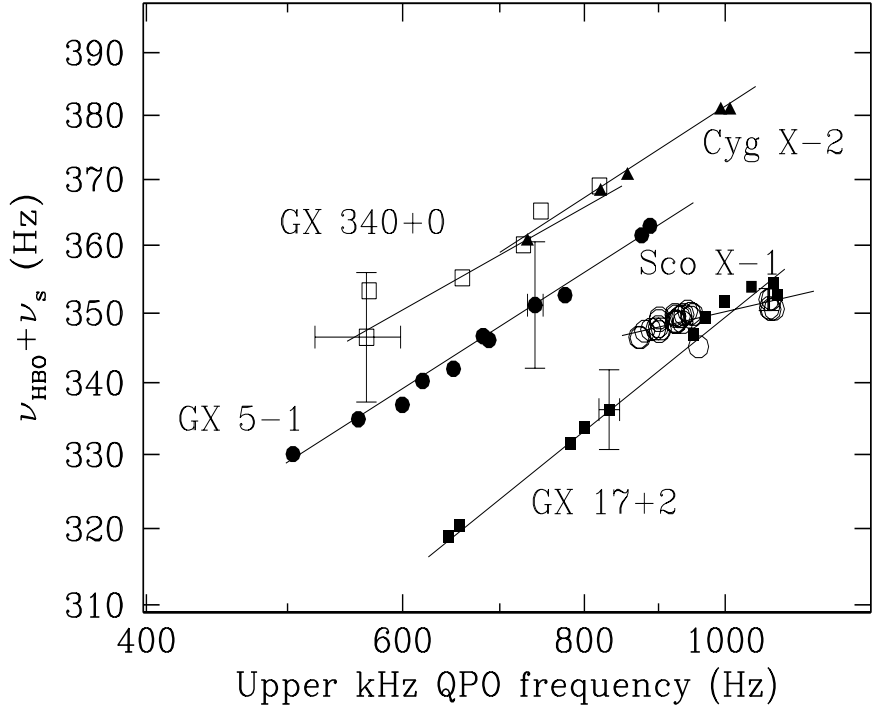,angle=-90,height=10truecm,width=12truecm}}
\figcaption[]
 {\small Correlations between $\nu_{\rm HBO}+\nu_s$,
the sum of the HBO frequency and the inferred spin
frequency, and the upper kilohertz QPO frequency in
five Z sources. Only a single typical error bar is
shown for each source, for clarity. The lines show the
best fits of the magnetospheric beat-frequency
relation~(\ref{magHBO}) to the data. The error bars
for the Sco~X-1 points are smaller than the circles
plotted. The error bars for the Cyg~X-2 points are
substantially larger than for the other points,
because of the large uncertainty in the Cyg~X-2 spin
frequency.} \end{figure}

In order to test the relations~(\ref{magHBO})
and~(\ref{nuindHBO}) predicted by the magnetospheric
beat-frequency model, simultaneous measurements of
$\nu_{\rm HBO}$, $\nu_{2}$, and $\nu_{s}$ are needed.
The HBO and kilohertz QPO frequencies are directly
observed, but oscillations at the neutron star spin
frequency have not yet been detected in the persistent
emission of any Z source. However, comparisons of the
frequencies of the two simultaneous kilohertz QPOs
observed in the persistent emission of the atoll
sources with the frequencies of the nearly coherent
oscillations observed in these sources during type~I
X-ray bursts indicate that the neutron star spin
frequency is nearly equal to the frequency separation
between the two kilohertz QPOs (Strohmayer et al.\
1996; Miller et al.\ 1998; Strohmayer et
al.\markcite{Setal98}\ 1998; Psaltis et
al.\markcite{Petal}\ 1998; Miller 1999). Hence, for
GX~17$+$2, GX~5$-$1, GX~340$+$0, and Cyg~X-2 we set
the spin frequency equal to the average frequency
separation of the kilohertz QPOs. In Sco~X-1, the
frequency separation of the kilohertz QPOs is
consistent with being constant at the lowest inferred
accretion rates but decreases at higher rates (van
der Klis et al.\ 1997). Sco~X-1 is thought to be
accreting at near-Eddington mass accretion rates when
the frequency separation of the kilohertz QPOs
decreases, and it is therefore plausible that this
decrease is related to the effects of radiation forces
on the dynamics of the accretion flow near the neutron
star (see, however, M\'endez et al.\markcite{Metal98}
1998; Psaltis et al.\ 1998). Hence, in plotting the
Sco~X-1 data, we set the spin frequency equal to the
nearly constant frequency separation of its kilohertz
QPOs at low inferred mass accretion rates. The spin
frequencies we have adopted are listed in Table~2.

Relation~(\ref{magHBO}) describes adequately ($\chi^2
\lesssim 1.5$ per degree of freedom) the dependence of
the sum of the HBO and inferred spin frequencies on
the frequency of the upper kilohertz QPO in the five Z
sources in our sample, considered separately. Table~2
lists the best-fit parameters and their $1\sigma$
errors and Figure~1 compares the best-fit relations
with the frequency data on each source. If $\lambda
\approx 1$, the power-law index $\alpha$ for all
sources except Sco~X-1 is $\approx 0.2$. This value is
consistent with the expectation that the Z sources are
accreting at near-Eddington rates and hence that the
inner accretion disk is optically thick and
radiation-pressure-dominated (see Table~1). For
Sco~X-1, however, a significantly weaker dependence of
the HBO frequency on $\dot{M}$ is required, if
$\lambda$ is independent of the mass accretion rate
(but see Appendix~A and Psaltis, Belloni, \& van der
Klis 1999).

\begin{figure}[t]
 \centerline{
\psfig{file=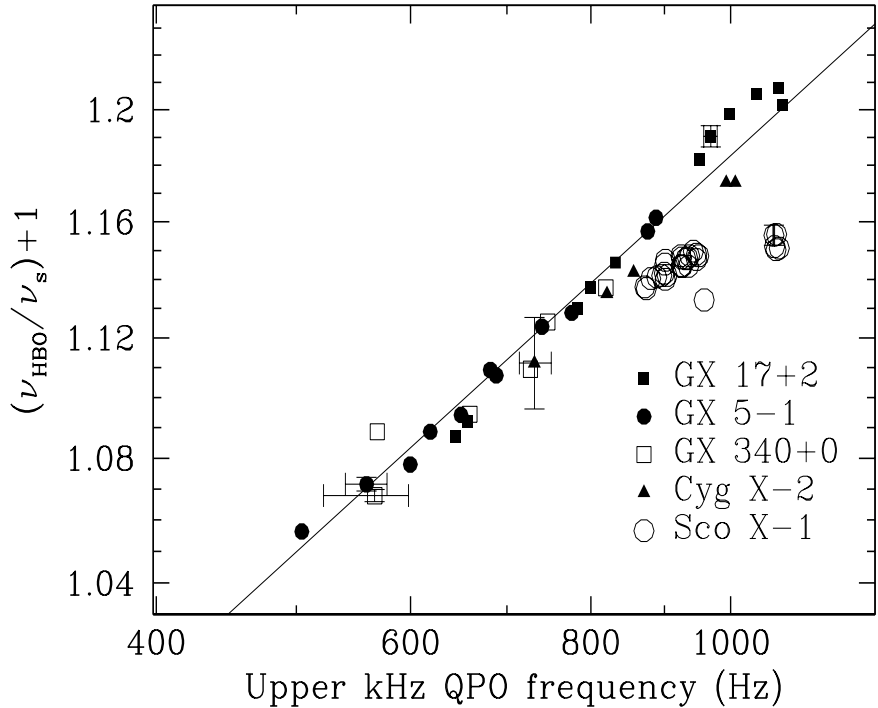,angle=-90,height=10truecm,width=12truecm}}
\figcaption[]
 {\small Correlation between $\nu_{\rm HBO}/\nu_s+1$
and the frequency of the upper kilohertz QPO in five Z
sources. The solid line is the best fit of the
magnetospheric beat-frequency
relation~(\ref{nuindHBO}) to the points with upper
kilohertz QPO frequencies less than 850~Hz.}
 \end{figure}

Figure~2 shows the quantity $\nu_{\rm HBO}/\nu_{s}+1$
plotted as a function of the upper kilohertz QPO
frequency $\nu_2$ for the five Z sources in our
sample. The frequency data on all the sources are
consistent with a single, universal relation between
$\nu_{\rm HBO}$, $\nu_{s}$, and $\nu_2$, as predicted
by equation~(\ref{nuindHBO}), when $\nu_2$ is $<
850$~Hz. This relation is shown as a solid line in
Figure~2. Figure~3 shows the confidence contours for
the power-law index $\alpha/\lambda$ and the
coefficient $A_2$ in relation~(\ref{nuindHBO})
obtained by fitting this relation to all the data with
$\nu_2 < 850$~Hz. Assuming that $\nu_0$ is $\approx
1200$~Hz, the best-fit value of $A_2$ gives a lower
bound on the critical fastness $\omega_c$, because
$\xi\dot{M}$ is expected to be a monotonically
increasing function of $\dot M$ and hence
$\langle(\xi\dot{M})^\alpha\rangle \le
(\xi\dot{M}_{\rm E})^\alpha$. If the magnetospheric
beat-frequency model is the correct explanation of the
HBO, then $\omega_c$ is $\gtrsim 0.8$ for the magnetic
field strengths and accretion rates of the Z sources.

\begin{figure}[t]
 \centerline{
\psfig{file=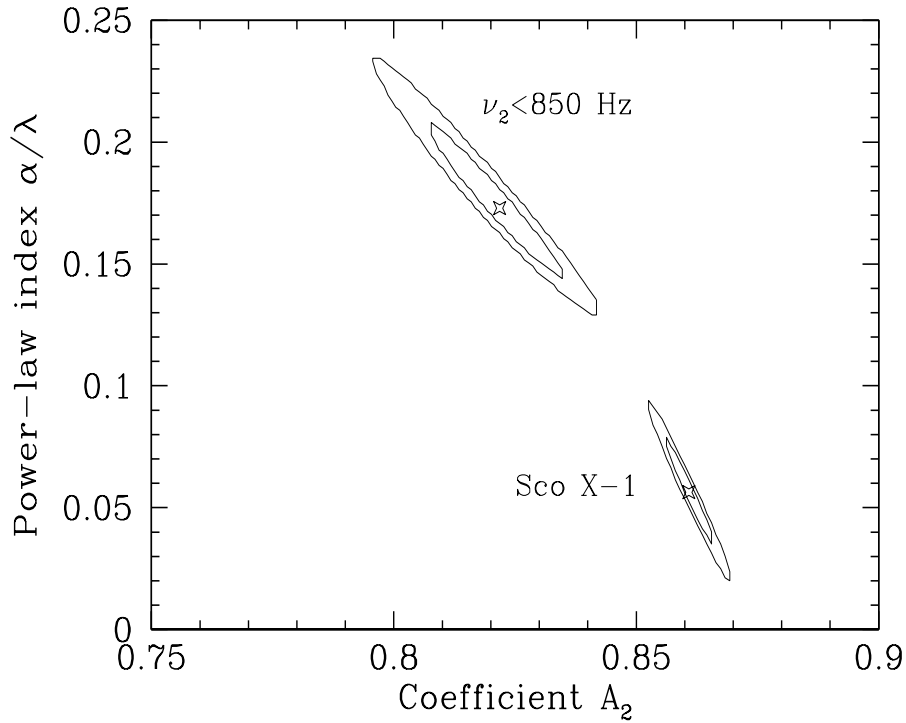,angle=-90,height=10truecm,width=12truecm}}
\figcaption[]
 {\small Confidence contours for the coefficient $A_2$
and the power-law index $\alpha/\lambda$ in the
beat-frequency relation~(\ref{nuindHBO}), obtained by
fitting this relation to the Z-source frequency data
with upper kilohertz QPO frequencies less than 850~Hz
and to the data of Sco~X-1. The inner and outer
contours show the 68\% and 99\% confidence limits,
respectively. The star indicates the best-fit values.
The best-fit relation is compared with the frequency
data in Fig.~2. Within the beat-frequency model of the
HBO, the value of $A_2$ is an upper bound on the
critical fastness $\omega_c$.}
 \end{figure}

When $\nu_2$ is $>850$~Hz, the HBO frequencies of
GX~17$+$2 are up to 2\% higher than predicted by
extrapolating the universal relation that holds at
lower frequencies, whereas those of Sco~X-1 are as
much as 5\% lower. This indicates that there is at
least one other important parameter that varies with
$\nu_2$. For example, the structure of the inner disk
may change at high accretion rates, causing the
exponent $\lambda$ to vary from source to source. This
conjecture cannot be tested without a specific model
for the variation in $\lambda$, because if $\lambda$
is chosen to reproduce the behavior of the data,
relation~(\ref{nuindHBO}) looses all predictive power.

The magnetospheric beat-frequency model of the HBO
requires that the neutron stars in all the Z sources
be near magnetic spin equilibrium. The tight, universal
correlation between the HBO, spin, and upper kilohertz
QPO frequencies in all the Z sources when
$\nu_2<850$~Hz is explained by the model if $A_2
\nu_0^{\alpha/\lambda}= \omega_c\nu_0^{a/\lambda}
{\langle(\xi\dot{M})^{\alpha}\rangle}/
{(\xi\dot{M}_{\rm E})^{\alpha}}$ is approximately the
same in all of them. All the Z sources are thought to
be accreting at very similar rates (comparable to the
Eddington critical accretion rate; see, e.g., Lamb
1989; Hasinger \& van der Klis 1989) and hence
${\langle(\xi\dot{M})^{\alpha}\rangle}/
{(\xi\dot{M}_{\rm E})^{\alpha}}$ is not expected to
differ much from one to another. The critical fastness
$\omega_c$ is expected to be comparable to unity and
may be a universal constant for a given inner disk
structure (Ghosh \& Lamb 1979, 1992). If so, then
$\nu_0^{\alpha/\lambda}$ is nearly the same in
the five Z sources in our sample and hence the
frequency of the upper kilohertz QPO is a good
absolute measure of the mass accretion rate in these
sources. Stated differently, {\it the relation between
the upper kilohertz QPO frequency and the accretion
rate appears to be very similar in all the Z sources}.

\subsection{Discussion}

The analysis presented in \S2.1 demonstrates that if
(a)~the neutron stars in the Z sources are spinning
near their magnetic spin equilibrium rates, (b)~the
frequency separation between the upper and lower
kilohertz QPOs is approximately equal to the
neutron-star spin frequency, (c)~the inner accretion
disk in the Z sources is optically thick and
radiation-pressure dominated, and (d)~the upper
kilohertz QPO frequency is proportional to the mass
accretion rate through the inner disk, then the
magnetospheric beat-frequency model is consistent with
the observed behavior of the HBO frequency and, in
particular, the tight, universal correlation
(Figure~2) between the HBO and kilohertz QPO
frequencies in all the Z sources in our sample. All of
these assumptions are expected to be satisfied, as
discussed in \S2.1.

As noted earlier, the magnetospheric beat-frequency
model of the HBO is consistent with models for the
upper kilohertz QPO that identify its frequency with
an orbital frequency in the inner disk only if a small
fraction of the accreting matter does not couple to
the stellar magnetic field at the radius $r_{m}$ but
instead remains in a geometrically thin Keplerian disk
down to the radius responsible for the upper kilohertz
QPO (i.e., $\xi$ must be less than unity; see van der
Klis 1998; Miller et al.\ 1998; and Alpar \& Yilmaz
1997 for discussions). In particular, the
interpretation of both the HBO and the two
simultaneous kilohertz QPOs as rotational beat
phenomena requires that there be two distinct radii in
the inner accretion disk at which beating of the
neutron star spin frequency with the orbital frequency
produces a QPO, as is the case, for example, in the
sonic-point model (Miller et al.\ 1998).

Assuming that the HBO is a magnetospheric
beat-frequency phenomenon, we can use the observed HBO
properties together with a general relation for the
coupling radius to constrain the magnetic dipole
moments of the neutron stars in the Z sources in a way
that is largely independent of the structure of the
inner accretion disk. In the Ghosh \& Lamb (1979)
model of disk-magnetosphere interaction, the radius
$r_{m}$ at which the stellar magnetic field strongly
couples to the gas in the accretion disk is given
implicitly by (see Ghosh \& Lamb 1991)
 \begin{eqnarray}
 r_{m} & \simeq &
 \left(\frac{B_\phi}{B_p}\right)^{2/7}
 \left(\frac{\Delta r}{r_{m}}\right)^{2/7} 
 \left(\frac{\mu^{4}}{G M
 M_\odot \xi^2\dot{M}^2}\right)^{1/7}
 \nonumber\\
 &\simeq& 3.3\times 10^6
 \left(\frac{B_\phi}{B_p}\right)^{2/7}
 \left(\frac{\Delta r}{r_{m}}\right)^{2/7}
 \xi^{-2/7}
 \mu_{27}^{4/7}M^{-1/7}
 \left(\frac{\dot{M}}{\dot{M}_{\rm 
    E}}\right)^{-2/7}~\mbox{cm}
 \label{r0}
 \end{eqnarray}
 for any model of the inner accretion disk. Here
$B_\phi/B_p$ is the mean azimuthal magnetic pitch in
the annulus of radial width $\Delta r/r_{m}$ in the
inner disk where the stellar field strongly interacts
with gas in the disk.  

If the stellar magnetic field is too weak, it cannot
couple strongly to the gas in the accretion flow well
above the stellar surface and hence cannot generate
magnetospheric beat-frequency oscillations. Hence in
the magnetospheric beat-frequency model, the coupling
radius $r_{m}$ must be larger than the neutron star
radius $R_{\rm NS}$, which requires
 \begin{equation}
 \mu_{27}\gtrsim 1.3\, \xi^{1/2} M^{1/4}
 \left(\frac{B_\phi}{B_p}\right)^{-1/2}
 \left(\frac{\Delta r/r_{m}}{0.01}\right)^{-1/2}
 \left(\frac{\dot{M}_{\rm max}}
            {\dot{M}_{\rm E}}\right)^{1/2}
 \left(\frac{R_{\rm
NS}}{10^6~\mbox{cm}}\right)^{7/4}\;,
 \label{mumin}
 \end{equation} 
 where $\dot{M}_{\rm max}$ is the maximum mass
accretion rate at which the HBO is detected. In
deriving inequality~(\ref{mumin}) we have neglected the
contributions of any higher multipole moments of the
stellar magnetic field that may be present near the
neutron star surface or may be induced by the
electrical currents flowing in the disk (see Psaltis,
Lamb, \& Zylstra 1996 for a discussion). For the
Keplerian frequency at the coupling radius to exceed
the neutron star spin frequency, which is also required
in the magnetospheric beat-frequency model (see Ghosh
\& Lamb 1979), the magnetic dipole moment must satisfy
 \begin{equation}
 \mu_{27}\lesssim 10\,M^{1/4}
 \left(\frac{B_\phi}{B_p}\right)^{-1/2}
 \left(\frac{\Delta r/r_{m}}{0.01}\right)^{-1/2}
 \left(\frac{\nu_{s}}{300~\mbox{Hz}}\right)^{7/6}\;,
 \label{mumax}
 \end{equation}
 where we have used the fact that $\xi \dot{M}\lesssim
\dot{M}_{\rm E}$. These upper and lower bounds on the
magnetic dipole moment depend only very weakly on the
neutron star mass. Figure~4 shows the resulting lower
(eq.~[\ref{mumin}]) and upper (eq.~[\ref{mumax}])
bounds on the magnetic dipole moments of the neutron
stars in the Z sources, as a function of the relative
width $\Delta r/r_{m}$ of the coupling region.

\begin{figure}[t]
 \centerline{
\psfig{file=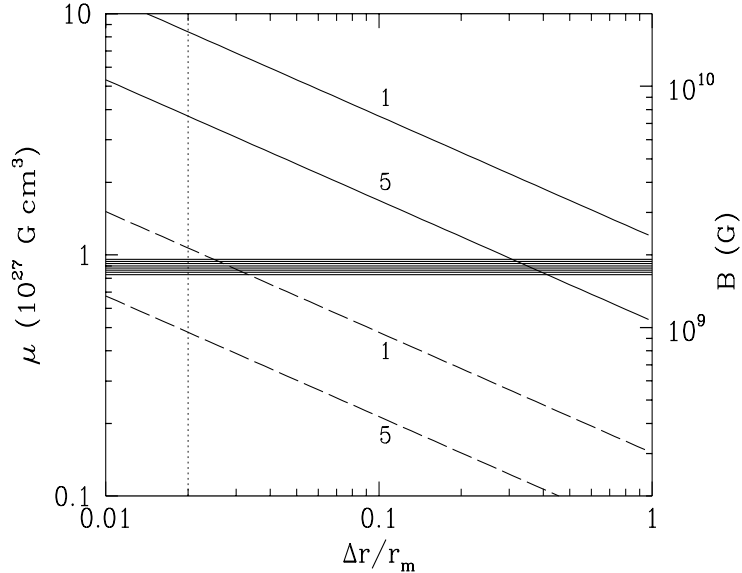,angle=-90,height=8truecm,width=10truecm}}
\figcaption[]
  {\small Bounds on the magnetic moments of the
neutron stars in the Z sources, derived from the
magnetospheric beat-frequency model of the HBO. {\it
Solid lines:\/} Upper bounds on the magnetic dipole
moment derived from the requirement that the orbital
frequency at the coupling radius be greater than the
neutron star spin frequency. {\it Dashed lines:\/}
Lower bounds on the magnetic dipole moment derived
from the requirement that the coupling radius be
larger than the radius of the star. Solid and dashed
lines are plotted for two values (1 and 5) of the mean
azimuthal magnetic pitch in the coupling region. {\it
Dotted line:\/} Lower bound on the fractional radial
width of the coupling region derived from the
requirement that the predicted FWHM of the HBO be
consistent with the observed width. {\it Shaded
region:\/} Range of dipole magnetic moments estimated
from the fits shown in Fig.~1 of the magnetospheric
beat-frequency relation~(\ref{magHBO}) to the Z-source
data. The polar magnetic field strength that
corresponds to a given magnetic moment is shown on the
right vertical axis. For simplicity, the neutron stars
in the Z sources were all assumed to have masses,
radii, and spin frequencies of $2M_\odot$, $10^6$~cm,
and 300~Hz.}
 \end{figure}

We can obtain an estimate of the magnetic dipole
moment of the Z sources by using the value of $A_1$
obtained by fitting relation~(\ref{magHBO}) to the
frequency data and the optically-thick,
radiation-pressure-dominated model of the inner disk.
The result is
 \begin{equation}
 \mu_{27}\simeq (0.8{\rm-}1.0)\; \xi^{0.3}
 \left(\frac{\nu_0}{1200~\mbox{Hz}}\right)^{-0.3}
 \left(\frac{M}{2M_\odot}\right)^{0.9}\;,
 \label{muest}
 \end{equation}
 where $\xi$ may depend on the magnetic field strength.
Relation~(\ref{muest}) shows that the HBO frequencies
predicted by the magnetospheric beat-frequency model
are consistent with the HBO frequencies observed if
$\mu_{27} \approx 1$, which implies that the dipole
magnetic fields of the Z sources have field strengths
at the magnetic poles of $\approx10^9$~G for a 10~km
neutron star. (Note that the estimated dipole magnetic
moment depends only weakly on the unknown parameters
$\xi$ and $\nu_0$.)

The relative width $\Delta r/r_{m}$ of the annulus
where the stellar magnetic field strongly couples to
the gas in the disk is expected to be greater than
$\sim 0.01$ (Ghosh \& Lamb 1992). Its value can be
bounded below using the observed FWHM of the HBO (see
also Alpar \& Shaham 1985; Lamb et al.\ 1985). 
Assuming that all other QPO broadening
mechanisms---such as lifetime broadening---are
negligible, we can estimate the relative width of the
annulus in the accretion disk in which the interaction
at the beat frequency affects the X-ray luminosity,
from the relative width of the HBO peak in power
spectra. The width of this annulus is necessarily
smaller than the width $\Delta r$ of the layer where
the magnetic field strongly interacts with the gas in
the disk, and hence
 \begin{equation}
 \frac{\Delta r}{r_{m}} \gtrsim \frac{2}{3}
 \left(\frac{\delta \nu_{\rm HBO}}{\nu_{\rm HBO}+
 \nu_{s}}\right)=
 0.02 \left(\frac{\delta \nu_{\rm HBO}}{10~\mbox{Hz}}
  \right)
 \left(\frac{350~\mbox{Hz}}{\nu_{\rm
HBO}+\nu_{s}}\right)\;,
 \label{width}
 \end{equation}
 where $\delta \nu_{\rm HBO}$ is the FWHM of the HBO. 
Figure~4 displays this bound on $\Delta r/r_{m}$ and
the constraint it imposes on the dipole moment of the
stellar magnetic field. 

Figure~4 shows that these additional physical bounds
on the dipolar magnetic fields of the Z sources
derived from the magnetospheric beat-frequency
interpretation of the HBO are consistent both with
each other and with the field strengths $\approx
10^9$~G estimated in equation~(\ref{muest}) and by
modeling the X-ray spectra of the Z sources (Psaltis
et al.\ 1995; Psaltis \& Lamb 1998).


\section{THE LENSE-THIRRING PRECESSION INTERPRETATION}

\subsection{Model Predictions and Comparison with 
Observations}

In the nodal (Lense-Thirring and tidal) precession
model of the HBO (Stella \& Vietri 1998), a narrow
ring or clumps of gas are assumed to be in a tilted
orbit at the radius responsible for the upper
kilohertz QPO and to precess with the frequency of a
test particle in such an orbit (Stella \& Vietri
1998). Alternatively, if the disk ends at this radius,
one of the many weakly damped global precession modes
localized near the inner edge of the disk (Markovi\'c
\& Lamb 1998) may be excited.

In the weak field limit, the nodal precession
frequency of an infinitesimally tilted orbit at the
radius where the frequency of a circular Keplerian
orbit is $\nu_{\rm K}$ is (see Stella \& Vietri 1998;
Morsink \& Stella 1999)
 \begin{eqnarray}
 \nu_{\rm NP}  &\approx  &13.2
\left(\frac{I_{45}}{M}\right)
 \left(\frac{\nu_{s}}{300~\mbox{Hz}}\right)
 \left(\frac{\nu_{\rm K}}
            {1~\mbox{kHz}}\right)^2\nonumber\\ &&
-4.7\left(\frac{I_{45}}{M^{5/3}}\right)
 \left(\frac{\eta}{0.01}\right)
 \left(\frac{\nu_{s}}{300~\mbox{Hz}}\right)^2
 \left(\frac{\nu_{\rm K}}
            {1~\mbox{kHz}}\right)^{7/3}\;,
 \label{clasfreq}  
 \end{eqnarray}
 where $\eta \equiv -(A/I_{zz})(\nu_{s}/300~{\rm
Hz})^{-2}$ in terms of $A$, the coefficient of the
quadrupole moment of the gravitational field, and
$I_{\rm zz}$, the neutron star moment of inertia with
respect to its spin axis. Equation~(\ref{clasfreq}) is
derived by expanding the full expression for $\nu_{\rm
NP}$ in a power series in $\nu_{s}$ and retaining only
terms up to second order.

{\it Lense-Thirring precession}.---If the effects of
the quadrupole component of the star's gravitational
field are negligible, the localized warping modes of
the inner disk will precess with a frequency close to
the Lense-Thirring frequency of a test particle
(Markovi\'c \& Lamb 1998), which is given by the first
term in equation~(\ref{clasfreq}). Identifying the
centroid frequency of the HBO with the Lense-Thirring
frequency at the radius where the orbital frequency is
equal to the frequency $\nu_2$ of the upper kilohertz
QPO gives (Stella \& Vietri 1998)
 \begin{equation}
 \nu_{\rm HBO}=\frac{8\pi^2 I}{c^2M} \nu_{s} 
  \nu_{2}^2
 = 13.2 \left(\frac{I_{45}}{M}\right)
 \left(\frac{\nu_{s}}{300~\mbox{Hz}}\right)
 \left(\frac{\nu_{2}}{1~\mbox{kHz}}
   \right)^2~~\mbox{Hz}\;,
 \label{LT}
 \end{equation}
 where $I\equiv 10^{45}I_{45}$~g cm$^2$ is the moment
of inertia of the neutron star and $\nu_{2}$ is the
orbital frequency at the radius responsible for the
upper kilohertz QPO. The X-ray visibility as well as
the excitation and damping of the precession modes of
the inner disk have not yet been addressed (see
Markovi\'c \& Lamb 1998 for a discussion).
Equation~(\ref{LT}) predicts a relation between the
HBO frequency, the spin frequency of the neutron star,
and the frequency $\nu_{2}$ of the upper kilohertz QPO
that depends only on the structure of the neutron
star, through the ratio $I/M$.

\begin{figure}[t]
 \centerline{
\psfig{file=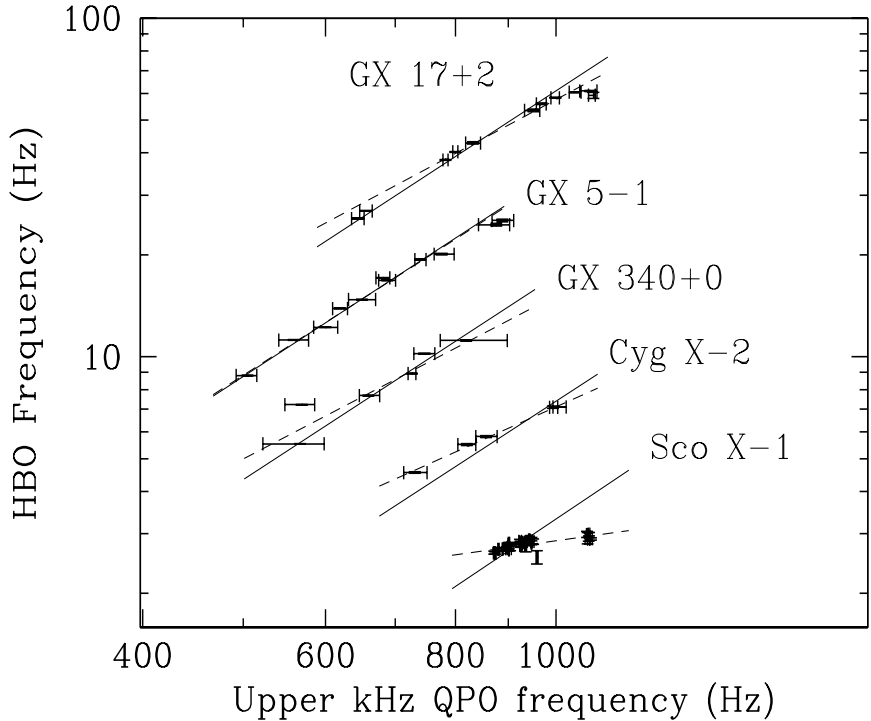,angle=-90,height=10truecm,width=12truecm}}
\figcaption[]
  {\small Correlations between HBO frequency and upper
kilohertz QPO frequency in five Z sources. The solid  
lines show the best fits of the Lense-Thirring
relation~(\ref{LT}) to the data. The dotted lines show
the best fits of the more general power-law
relation~(\ref{fitlog}), which is discussed in
Appendix A. The HBO frequencies and lines for all the
sources except GX~17$+$2 have been shifted downward by
successive factors of two for clarity.}
 \end{figure}

Figure~5 shows the HBO frequencies observed in the
five Z sources in the present sample, plotted against
the frequencies of their upper kilohertz QPOs.
Separate fits of equation~(\ref{LT}) to the data on
the individual sources, using the neutron star spin
frequency inferred from the frequency separation of
the kilohertz QPOs and treating $I/M$ as a free
parameter, give values of $\chi^2$ per degree of
freedom of order unity for four of the Z sources but
$\sim 4$ for Sco~X-1. There is no other freedom in
relation~(\ref{LT}) and hence pure Lense-Thirring
precession must be rejected as an explanation of the
HBO. Moreover, as Stella \& Vietri (1998) noticed (see
also Wijnands et al.\ 1998a; Jonker et al.\ 1998), the
coefficients of $\nu_{s} \nu_{2}^2$ required to fit
the data give values of $I_{45}/M \gtrsim 4$, which
is $\sim 4$--5 times larger than the largest ratios
given by realistic equations of state for stars of any
mass and about 2.5 times larger even than the largest
ratio given by the extremely stiff relativistic
mean-field equation of state L (see Table~3).

\begin{figure}[t]
 \centerline{
\psfig{file=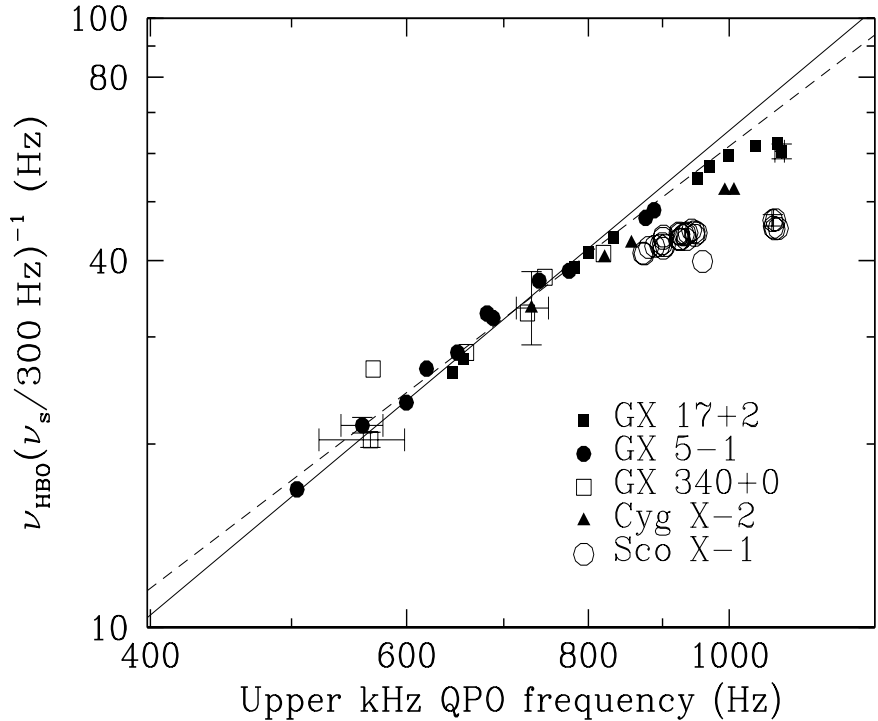,angle=-90,height=10truecm,width=12truecm}}
\figcaption[]
 {\small Correlation between the HBO frequency
$\nu_{\rm HBO}$ divided by the inferred neutron star
spin frequency $\nu_{s}$ (Table~2) and the upper
kilohertz QPO frequency in five Z sources. Only a
single typical error bar is shown for each source, for
clarity. The solid line is the best fit of the
Lense-Thirring relation~(\ref{LT}) to all points with
upper kilohertz QPO frequencies less than 850~Hz. The
dotted line is the best fit of the more general
power-law relation~(\ref{fit}) discussed in Appendix~A
to the same points.}
 \end{figure}

In the Lense-Thirring precession model of the HBO, the
relation between $\nu_{\rm HBO}/\nu_{s}$ and the upper
kilohertz QPO frequency depends only on the mass of
the star and the equation of state of neutron-star
matter (see eq.~[\ref{LT}]). Data from similar neutron
stars should therefore follow similar relations.
Figure~6 shows how the ratio $\nu_{\rm HBO}/\nu_{s}$
scales with the frequency $\nu_2$ of the upper
kilohertz QPO; according to the Lense-Thirring
precession model, this ratio should scale as
$\nu_2^2$. Figure~6 shows that the data with
frequencies $\nu_2<850$~Hz are consistent
($\chi^2_{\rm d.o.f} \approx 1.1$) with a single
relation of the form~(\ref{LT}). Again, however, the
coefficient given by the fit requires neutron stars
with $I_{45}/M \gtrsim 4$, which is implausibly
large. Furthermore, the points that have
$\nu_2>850$~Hz are inconsistent with the relation of
the form~(\ref{LT}) that fits the points with lower
values of $\nu_2$.

{\it Effect of classical precession}.---Stella \&
Vietri (1998; see also the extended discussion in
Morsink \& Stella 1999) suggested that the flattening
of the $\nu_{\rm HBO}$--$\nu_2$ correlation at high
$\nu_2$ might be caused by the increasing importance,
as the disk penetrates closer to the star, of the
classical precession caused the rotation-induced
quadrupole component of the star's gravitational
field. This precession is retrograde but smaller than
the prograde gravitomagnetic precession and therefore
tends to reduce the nodal precession frequency. We can
test this suggestion quantitatively using the data for
Sco~X-1 and GX~17$+$2, which deviate most strongly
from the relation of the form~(\ref{LT}) that fits the
points with low values $\nu_2<850$~Hz..

\begin{figure}[t]
 \centerline{
\psfig{file=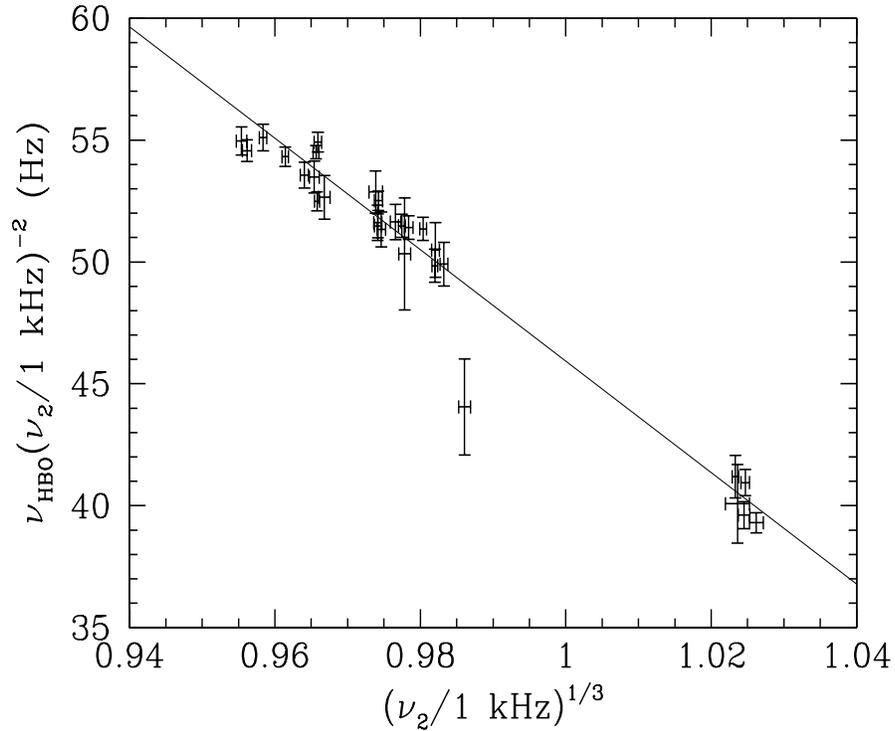,angle=-90,height=10truecm,width=12truecm}}
\figcaption[]
 {\small Correlation in Sco~X-1 between $\nu_{\rm
HBO}/\nu_2^2$ and $\nu_2^{1/3}$, where $\nu_{\rm HBO}$
is the HBO frequency and $\nu_2$ is the frequency of
the upper kilohertz QPO. The solid line is the
straight line that best fits the data.}
 \end{figure}

If the HBO is caused by nodal precession and classical
precession is important, then $\nu_{\rm HBO}/\nu_2^2$
should decrease linearly with increasing $\nu_2^{1/3}$
(cf.\ Stella \& Vietri 1998), because
relation~(\ref{clasfreq}) can be rewritten as
 \begin{eqnarray}
 \left(\frac{\nu_{\rm HBO}}{1~\mbox{Hz}}\right)
 \left(\frac{\nu_2}{1~\mbox{kHz}}\right)^{-2} & =
&13.2 \left(\frac{I_{45}}{M}\right)
 \left(\frac{\nu_{s}}{300~\mbox{Hz}}\right)\nonumber\\
& & -4.7\left(\frac{I_{45}}{M^{5/3}}\right)  
 \left(\frac{\eta}{0.01}\right)
 \left(\frac{\nu_{s}}{300~\mbox{Hz}}\right)^2
 \left(\frac{\nu_2}{1~\mbox{kHz}}\right)^{1/3}\;.
 \label{clas}
 \end{eqnarray}
 Figure~7 plots $\nu_{\rm HBO}/\nu_2^2$ against
$\nu_2^{1/3}$ for Sco~X-1 and also shows the best-fit
straight line with slope $1/3$, which has $\chi^2_{\rm
d.o.f.} \approx 1$. The fact that the data in Figure~7
can be fit satisfactorily by a straight line with
slope $1/3$ is not strong evidence for this scaling,
because the range of measured $\nu_2^{1/3}$ values is
very narrow. However, we can use the best-fit value of
the intercept of the straight line with the vertical
axis to estimate the value of the parameter $\eta$ that
characterizes the quadrupole moment of the
gravitational field and to estimate $I/M$. The results
are
 \begin{equation}
 \eta_{\rm Sco}
 \simeq 2.3\times 10^{-2} M^{2/3}
 \left(\frac{\nu_{s}}{300~\mbox{Hz}}\right)^{-1}
 \end{equation} and
 \begin{equation}
 \left(\frac{I_{45}}{M}\right)
 \left(\frac{\nu_{s}}{300~\mbox{Hz}}\right)
 =20.8 \pm 2.1\;.
 \label{InuM}
 \end{equation}
 The value of $I/M$ required by equation~(\ref{InuM})
is $\sim 20$ time larger than the largest values given
by realistic neutron-star equations of state (see
Table~3). The deviation of the GX~17$+$2 data from the
power-law relation~(\ref{LT}) at high frequencies (see
Fig.~6) requires that the classical precession
frequency be negligible for $\nu_2<850$~Hz but
comparable to the Lense-Thirring precession frequency
at slightly larger values of $\nu_2$. This is not
possible, because the exponents of $\nu_2$ in the
Lense-Thirring and classical precession terms of
equation~(\ref{clasfreq}) are too similar. The data
are therefore inconsistent with the predictions of the
simple nodal precession model.

\subsection{Discussion}

In the Lense-Thirring precession model of the HBO, the
HBO frequency, the spin frequency of the neutron star,
and the frequency of the upper kilohertz QPO are
related by equation~(\ref{LT}). As discussed in the
previous section, the data are consistent with this
relation when the frequency of the upper kilohertz QPO
is $<850$~Hz, but the inferred value of $I/M$ is
implausibly large (see also Stella \& Vietri 1997,
1998; Morsink \& Stella 1999). Aside from the rather
unlikely possibility that neutron stars have ratios of
$I/M$ that are four times as large as the largest
values for stellar models constructed with realistic
equations of state, three other possibilities have
been suggested for reducing this large discrepancy.

First, the frequency difference $\Delta\nu$ between the
kilohertz QPOs might be equal to half the neutron star
spin frequency $\nu_s$ rather than equal to it. This is
very unlikely in any beat-frequency model of the
kilohertz QPOs, because it would require a special
direction that rotates with the neutron star but
affects the inner accretion disk only once every two
beat periods. However, $\Delta\nu=\frac{1}{2}\nu_s$
appeared possible, given the initial analysis of the
data taken during the type~I X-ray bursts of
4U~1636$-$536 (Zhang et al.\markcite{Zetal97} 1997;
Strohmayer et al.\markcite{Setal98} 1998), which
showed a strong oscillation at about 580~Hz,
approximately twice the frequency separation of the
two kilohertz QPOs. However, further analysis of this
data by Miller (1999) using a matched-waveform
filtering technique has revealed the presence of a weak
coherent oscillation at about 290~Hz, approximately
equal to the frequency separation of the two kilohertz
QPOs. Thus, it now appears very unlikely that the spin
frequencies of these neutron stars are twice the
frequency separation of their kilohertz QPOs.

Second, the observed HBO frequencies and their second
harmonics might represent the second and fourth
harmonics of the fundamental Lense-Thirring
frequency~(\ref{LT}), rather than the first and second
harmonics. Indeed, a precessing circular orbit has a
two-fold symmetry that could, in principle, produce
even-order harmonics that are stronger than the
odd-order harmonics. Moreover, power-density spectra
of the Z sources show a relatively strong, broad-band
noise component at frequencies comparable to the ones
predicted by relation~(\ref{LT}). This so-called
low-frequency noise (Hasinger \& van der Klis 1989)
might inhibit detection of the fundamental of a
low-frequency precession frequency (see, e.g.,
Fig.\,6a of Kuulkers et al.\markcite{Ketal94} [1994]
for peaked features in the low-frequency noise
component of GX~5$-$1).  Determination of the upper
limits on the amplitudes of any QPOs at these
frequencies would significantly constrain this
possibility. Note, however, that even if the HBO and
its overtone are the second and fourth harmonics of
the precession frequency, the $I/M$ ratios required to
explain the HBO observations would still be a factor
$\gtrsim 2$ larger than predicted by realistic
neutron-star equations of state.

A further difficulty with the Lense-Thirring precession
interpretation of the HBO is that the observed
correlation between the HBO and kilohertz QPO
frequencies is significantly different from what
is predicted by relation~(\ref{LT}) when the frequency
of the upper kilohertz QPO is $>850$~Hz. As
demonstrated in \S3.1, this difference cannot be
explained by classical precession; nor can it be
explained by strong-field corrections to
relation~(\ref{LT}) (Stella \& Vietri 1998). 

A third possibility is that radiation forces increase
the ratio of $\nu_{\rm NP}$ to $\nu_2$ by the factor
$\sim2$--5 required to bring it into agreement with
the observed HBO and upper kHz QPO frequencies. The Z
sources are thought to be accreting at near-Eddington
mass accretion rates when the frequencies of the upper
kilohertz QPOs are comparable to $\sim 1$~kHz (see,
e.g., Psaltis et al.\ 1995). Hence radiation forces,
which were neglected in equation~(\ref{clasfreq}),
almost certainly are important. At near-critical
luminosities, both orbital and nodal precession
frequencies can be altered by large factors compared
to their values in the absence of radiation; hence
radiation forces might possibly explain the large
discrepancy between the observed frequencies of the
HBO and the frequencies predicted by the nodal
precession model.

An explanation in terms of the combined effects of
Lense-Thirring precession and radiation forces would,
however, require the physically implausible result
that radiation forces leave the variation with
$\nu_{2}$ basically unchanged while increasing the
ratio of $\nu_{\rm NP}$ to $\nu_2$ by a factor
$\sim2$--5. Such an explanation would also require
that the QPO peaks not be significantly broadened by
the radiation drag force at the same time that
radiation forces are strong enough to change the
orbital and precession frequencies by a factor
$\sim2$--5. The HBO peaks in Sco~X-1, for example,
have fractional widths $\delta\nu/\nu\lesssim 0.5$
even when the inferred accretion rate is near the
Eddington critical rate (van der Klis et al.\ 1997).
More fundamentally, if radiation forces do change the
orbital and precession frequencies of gas accreting
onto the Z sources by large factors, as they may well
do, the observed correlation between the HBO and upper
kilohertz QPO frequencies would be explained primarily
by the effect of the radiation forces and not by the
gravitomagnetic torque.


\section{CONCLUSIONS}

In \S2 and \S3 we have studied in detail the behavior
of the HBO frequencies observed in five Z sources and
in particular their correlation with the frequencies
of the kilohertz QPOs, comparing the observed behavior
with the behaviors predicted by the magnetospheric
beat-frequency (Alpar \& Shaham 1985; Lamb et al.\
1985) and Lense-Thirring precession (Stella \& Vietri
1998) models of the HBO.

In \S2 we showed that the magnetospheric
beat-frequency model is consistent with the observed
correlation between the HBO and upper kilohertz QPO
frequencies in the five Z sources studied here if, as
expected, the neutron stars in these sources are
spinning near their magnetic spin equilibrium rates,
the frequency separation between the upper and lower
kilohertz QPOs is approximately equal to the
neutron-star spin frequency, the inner part of their
accretion disks are optically thick and
radiation-pressure-dominated, and the frequency of the
upper kilohertz QPO is approximately proportional to
the mass accretion rate. The model predicts a
universal relation between the horizontal branch
oscillation, stellar spin, and upper kilohertz QPO
frequencies that agrees well with the data on five Z
sources. The spin rates predicted by the model are
consistent with the range of the spin frequencies of
the Z sources inferred from the frequency separation
of their kilohertz QPOs if they are all accreting at
similar, near-critical rates and all have
$10^9$--$10^{10}$~G dipole magnetic fields. Such
magnetic fields are consistent with models of Z-source
X-ray spectra. The inferred value of the critical
fastness for the accretion rates and magnetic field
strengths of the Z sources is $\gtrsim 0.8$. If the
frequency of the upper kilohertz QPO is an orbital
frequency in the accretion disk, the magnetospheric
beat-frequency model requires that a fraction of the
accreting gas does not couple strongly to the stellar
magnetic field until it has penetrated to within a few
kilometers of the neutron star surface.

In \S3, we showed that the trend of the correlation
between the HBO frequency and the upper kilohertz QPO
frequency observed at upper kilohertz QPO frequencies
$\nu_2<850$~Hz agrees with the trend predicted by the
Lense-Thirring precession model. However, the observed
trend is inconsistent with the model for
$\nu_2>850$~Hz. The observed magnitudes of the HBO
frequencies are $\gtrsim 4$--5 times smaller than the
magnitudes predicted by the Lense-Thirring precession
model for realistic neutron-star equations of state.
Thus, in order to be consistent with the observed
magnitudes, either $I/M$ must be $\gtrsim 4$--5 times
larger than expected or the principal frequency of the
X-ray oscillation generated by nodal precession must
be $\gtrsim 4$--5 times the nodal precession frequency.

\acknowledgements

We thank Greg Cook for making available the numerical
code used to compute the suite of neutron star models
used in this work. DP acknowledges Charles Gammie for
many useful discussions on the physics of warped
accretion disks. DP also thanks L.\ Stella, V.\
Kalogera, R.\ Narayan, A.\ Esin, C.\ Gammie, K.\
Menou, and E.\ Quataert for discussions on the
observational evidence for the Lense-Thirring effect.
This work was supported in part by a post-doctoral
fellowship of the Smithsonian Institute (DP), by the
Netherlands Foundation for Research in Astronomy
(ASTRON) grant 781-76-017 (RW, JH, PJ, MvdK), by NSF
grant AST~96-18524 (FKL), by NASA grants NAG~5-2925
(FKL), NAG~5-2868 (MCM), NAG~5-3269 and NAG~5-3271
(JvP), and by several {\it RXTE\/} observing grants.
WHGL gratefully acknowledges support from NASA. MvdK 
gratefully acknowledges the Visiting Miller Professor
Program of the Miller Institute for Basic Research in
Science (UCB).

\appendix

\begin{center}
{\bf APPENDIX}
\end{center}

 \section{AN EMPIRICAL DESCRIPTION OF THE CORRELATION
OBSERVED BETWEEN THE HBO AND UPPER KILOHERTZ QPO
FREQUENCIES}

In this appendix we show that the correlations observed
between the HBO frequency $\nu_{\rm HBO}$, the upper
kilohertz QPO frequency $\nu_2$, and the frequency
separation $\Delta\nu$ between the two kilohertz QPOs
can be characterized by simple power-law relations
among the frequencies involved. Our purpose is to
facilitate comparison of the present data with future
data (see, e.g., Psaltis et al.\ 1999) or other
theoretical models.

\begin{figure}[t]
 \centerline{
\psfig{file=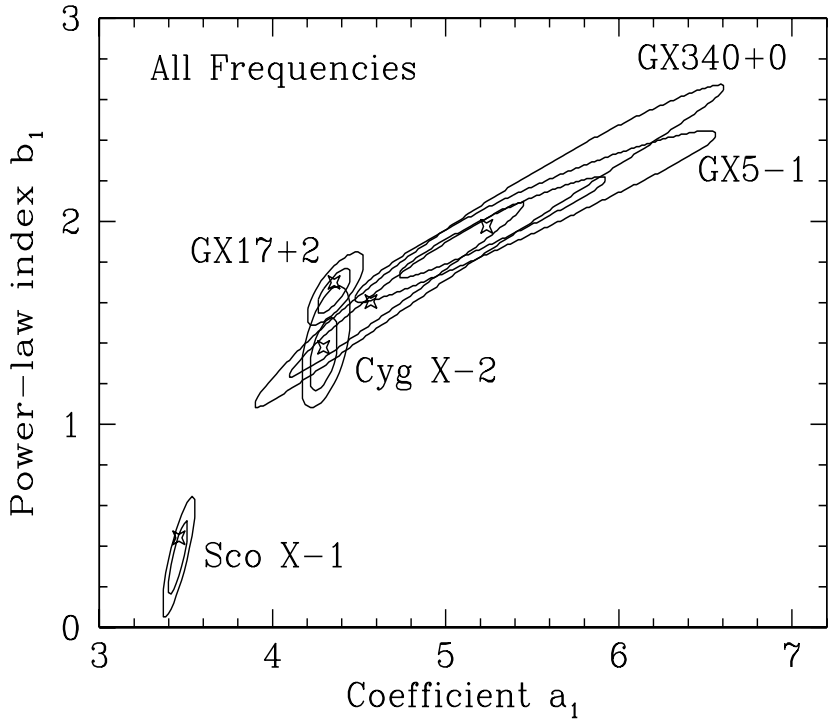,angle=-90,height=10truecm,width=12truecm}}
\figcaption[]
 {\small Confidence contours for the coefficient $a_1$
and the power-law index $b_1$, obtained by fitting
equation~(\ref{fitlog}) to the frequency data on each
of five Z sources, considered individually. The inner
and outer contours show the 68\% and 99\% confidence
limits, respectively, while the stars indicate the
best-fit values for each source. The best-fit
relations are shown as dashed lines in Fig.5.}
 \end{figure}

\begin{figure}[t]
 \centerline{
\psfig{file=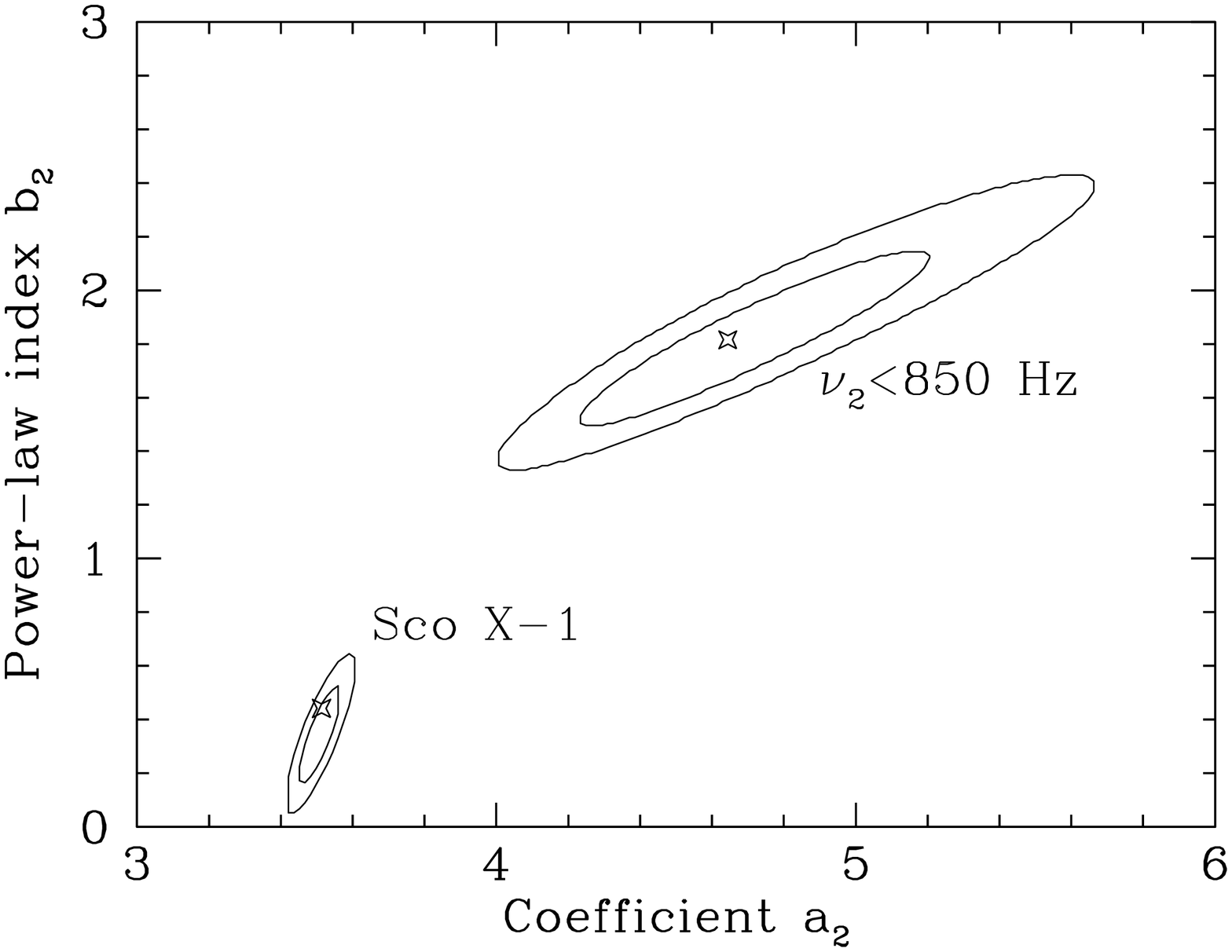,angle=-90,height=10truecm,width=12truecm}}
\figcaption[]
 {\small Confidence contours for the coefficients
$a_2$ and power-law indices $b_2$ derived by fitting
relation~(\ref{fit}) to two disjoint sets of frequency
data. The contours marked ``Sco~X-1'' were obtained by
fitting this relation to the Sco~X-1 data. The
contours marked ``$\nu_2 < 850$~Hz'' were obtained by
fitting this relation to the frequency data on all the
other sources, when the upper kilohertz QPO frequency
is less than 850~Hz; the best-fit version of this
relation is shown by the dashed line in Fig.~6. The
inner and outer contours show the 68\% and 99\%
confidence limits, respectively, while the stars
indicate the best-fit values for each source.}
 \end{figure}

The frequency correlations in the five Z sources in
our sample can be described adequately by the
power-law relation
 \begin{equation}
 \nu_{\rm HBO}= 13.2\, a_1\;
 \left(\frac{\nu_2}{1~\mbox{kHz}}\right)^{b_1}
 \;{\rm Hz}\;,
 \label{fitlog}
 \end{equation}
 where the constants $a_1$ and $b_1$ are different for
each source. The confidence contours obtained by the
fitting this relation to the measured HBO and upper
kilohertz QPO frequencies of the five Z sources in the
present sample are shown in Figure~8. The power-law
index that describes the Sco~X-1 data is significantly
smaller than the index that describes the data on the
other four sources. The best-fit relations for each
source are the dashed lines shown in Figure~5.

\begin{figure}[t]
\begin{minipage}[b]{10.0cm}
 \centerline{
\psfig{file=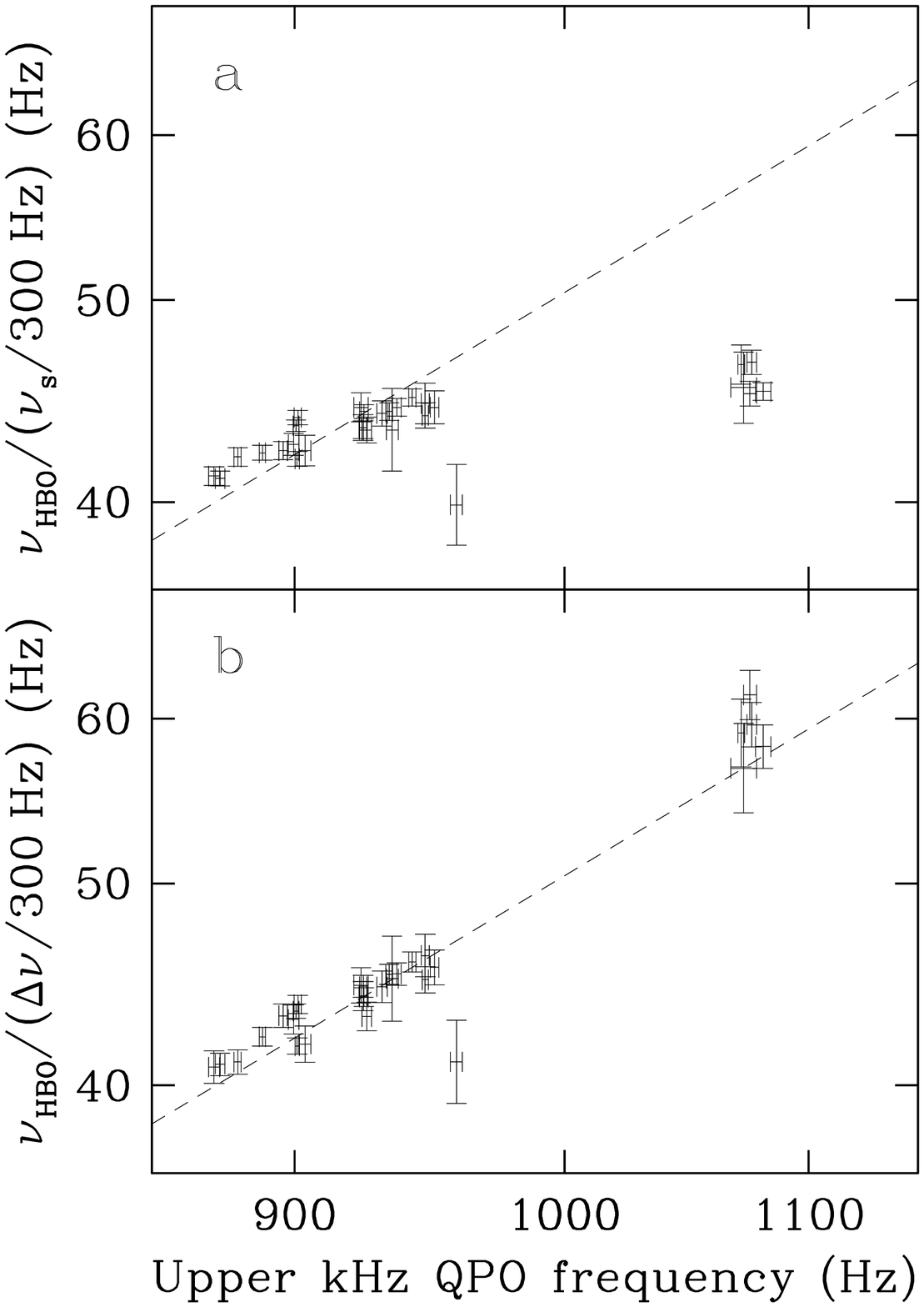,angle=0,height=12truecm,width=9truecm}}
\end{minipage}
\begin{minipage}[b]{5.9cm}
\figcaption[]
 {\small {\it Top panel\/}: Correlation observed in
Sco~X-1 between $\nu_{\rm HBO}/\nu_s$, the HBO
frequency divided by the neutron star spin frequency
inferred from the frequency separation of the two
kilohertz QPOs at low countrates, and the frequency of
the upper kilohertz QPO.
 {\it Bottom panel\/}: Correlation between $\nu_{\rm
HBO}/\Delta\nu$, the HBO frequency divided by the
instantaneous frequency separation of the kilohertz
QPOs, and the frequency of the upper kilohertz QPO.
In both panels the dashed line shows the best fit of
relation~(\ref{relation}) to the data.}
 \vspace*{1.8cm}
 \end{minipage}
 \end{figure}

For all the Z sources except Sco~X-1, the best-fit
value of the parameter $a_1$ is approximately
proportional to the spin frequency inferred from the
frequency separation of the two kilohertz QPOs (see
Fig.~6). Indeed, when the upper kilohertz QPO
frequency is $<850$~Hz, the correlation between
$\nu_{\rm HBO}$, $\nu_{s}$, and $\nu_2$ is described
adequately by the relation
 \begin{equation}
 \nu_{\rm HBO} = 13.2\, a_2 
 \left(\frac{\nu_{s}}{300~\mbox{Hz}}\right)
 \left(\frac{\nu_2}{1~\mbox{kHz}}\right)^{b_2}\;,
 \label{fit}
 \end{equation}
 with $a_2 \approx 4.6$, and $b_2 \approx 1.8$. The
confidence contours obtained by fitting this relation
to the Sco~X-1 points and to the points on the other
four sources for which $\nu_2<850$~Hz are shown in
Figure~9. As Figure~6 shows, the frequency correlation
is significantly flatter when $\nu_2$ is $>850$~Hz.

\begin{figure}[t]
 \centerline{
\psfig{file=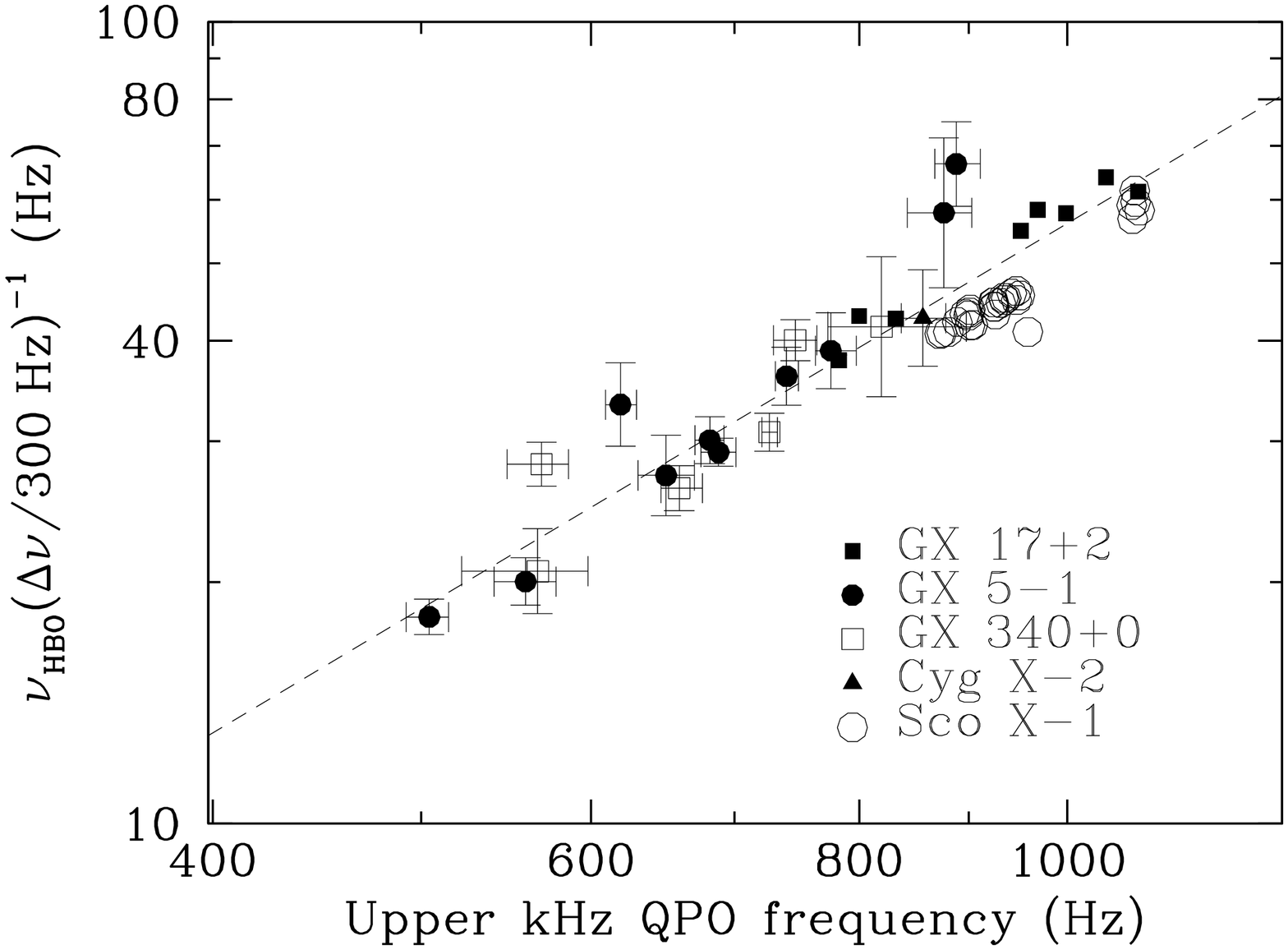,angle=-90,height=10truecm,width=12truecm}}
\figcaption[]
 {\small Correlation between the HBO frequency
$\nu_{\rm HBO}$ divided by the instantaneous frequency
separation $\Delta\nu$ of the kilohertz QPOs and the
frequency of the upper kilohertz QPO in five Z
sources. The uncertainties in the GX~17$+$2 and
Sco~X-1 points are typically smaller than the sizes of
the symbols plotted for these sources and have
therefore been omitted. The dashed line is the best
fit of relation~(\ref{relation}) to the frequency data
on all the sources except Sco~X-1.}
 \end{figure}

\begin{figure}[t]
 \centerline{
\psfig{file=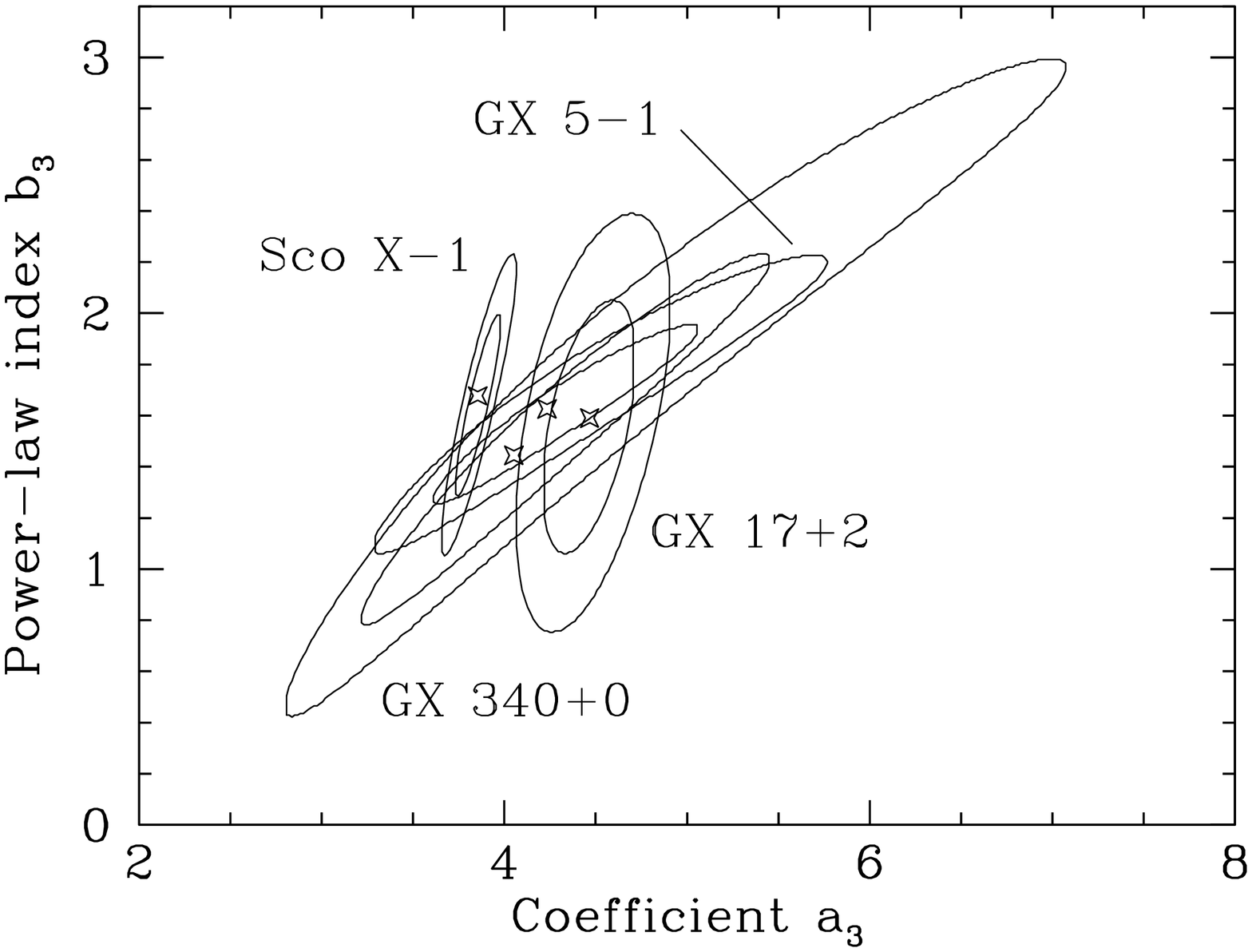,angle=-90,height=10truecm,width=12truecm}}
\figcaption[]
 {\small Confidence contours for the coefficients
$a_3$ and power-law indices $b_3$ derived by fitting
relation~(\ref{relation}) separately to the frequency
data on four of the five Z sources in our sample.
Cyg~X-2 was not considered because there is only one
data point for it. The inner and outer contours show
the 68\% and 99\% confidence limits, respectively,
while the stars indicate the best-fit values for each
source.}
 \end{figure}

In Sco~X-1, the HBO and kilohertz QPOs were
simultaneously detected mostly when it was on the
normal branch. In the other four Z sources, the HBO
and kilohertz QPOs were simultaneously detected mostly
when they were on their horizontal branches. The
transition from the horizontal to the normal branch is
thought to take place when the mass accretion rate
increases to within a few percent of the Eddington
critical rate (see Lamb 1989; Psaltis et al.\ 1995).
If so, the resulting change in the accretion flow
pattern (see Lamb 1989) might be responsible for the
different dependences of the HBO frequency $\nu_{\rm
HBO}$ and the instantaneous frequency separation
$\Delta \nu$ of the two kilohertz QPOs on the upper
kilohertz QPO frequency $\nu_2$ in Sco~X-1, compared
to the dependences in the other four Z sources in our
sample.

The ratio of $\nu_{\rm HBO}$ to $\Delta \nu$ in
Sco~X-1 increases more steeply with $\nu_2$ than does
the ratio of $\nu_{\rm HBO}$ to the (constant)
inferred spin frequency $\nu_{s}$. This is
demonstrated most clearly by the correlation plots
shown in Figure~10. (For all the Z sources in the
sample except Sco~X-1, $\Delta \nu$ is consistent with
being constant or with varying in the same way as it
does in Sco~X-1 [Psaltis et al.\ 1998; see also
Wijnands et al.\ 1997, 1998b; Jonker et al.\ 1998].
This is true mostly because of the relatively large
uncertainties in the measured kilohertz QPO
frequencies.) The dependence of $\nu_{\rm HBO}/\Delta
\nu$ on $\nu_2$ in Sco~X-1 is more consistent with the
behavior seen in the other four Z sources and suggests
that we plot $\nu_{\rm HBO}/\Delta \nu$ against
$\nu_2$ for all five of the Z sources in our sample.
The result is shown in Figure~11. In this plot we have
included only points derived from simultaneous
observations of the HBO and kilohertz QPO frequencies.
The larger scatter of the points in Figure~11 at
$\nu_2<850$~Hz compared to the scatter of the points
in Figure~7 corresponding to the same values of
$\nu_{\rm HBO}$ and $\nu_2$ is caused by the large
uncertainties in $\Delta \nu$. The points plotted in
Figure~11 are consistent with power-law relations of
the form
 \begin{equation}
 \nu_{\rm HBO} = 13.2\, a_3
 \left(\frac{\Delta \nu}{300~\mbox{Hz}}\right)
 \left(\frac{\nu_2}{1~\mbox{kHz}}\right)^{b_3}\;.
 \label{relation}
 \end{equation}
 The confidence contours obtained by fitting this
relation to the data on each of the sources except
Cyg~X-2 are shown in Figure~12. Cyg~X-2 was not
included because its HBO and kilohertz QPO frequencies
were measured simultaneously only once. These contours
show that a single universal relation of the
form~(\ref{relation}) with $a_3=4.2$ and $b_3=1.6$ is
consistent with the data on all the sources except
Sco~X-1. In fact, $b_3=1.6$ is consistent with all the
data. The Sco~X-1 and GX~17$+$2 data have relatively
small uncertainties and give best-fit coefficients
$a_3$ that are slightly but significantly different.
The uncertainties in the data on GX~5$-$1 and
GX~340$+$0 are sufficiently large that their contours
allow values of $a_3$ that are consistent with either
the Sco~X-1 or the GX~17$+$2 value.

The surprising correlation between the HBO frequency,
the frequency of the upper kilohertz QPO, and the {\it
instantaneous\/} frequency separation of the kilohertz
QPOs shown in Figure~11 may be coincidental. The
relatively large uncertainties in the currently
measured kilohertz QPO frequencies in all the Z
sources except Sco~X-1 prevent us from drawing any
firm conclusions about the significance of this
correlation. However, the strikingly similar relation
between the lower and upper kilohertz QPO frequencies
in all LMXBs that show kilohertz QPOs (Psaltis et al.\
1998) together with the correlation shown in Figure~11
suggests that the varying frequency separation of the
kilohertz QPOs in Sco~X-1 is a general property of the
kilohertz QPOs and is related, directly or indirectly,
to the frequency of the HBO. Additional data are
needed to test directly this conjecture (see Psaltis
et al.\ 1999 for an alternative possibility).


{\small 
 \begin{deluxetable}{ccccc}
 \tablewidth{350pt}
 \tablecaption{Scaling of the Keplerian Frequency at
the Coupling Radius\tablenotemark{a}} \tablehead{Disk
Model\tablenotemark{b} & $\nu_{\rm K,0}$~(Hz) &
$\alpha$ & $\beta$ & $\gamma$}
 \startdata
 1G & 430 & 0.38 & $-$0.87 & 0.82\\
 1R & 210 & 0.23 & $-$0.77 & 0.70\\
 2B & 80 & 0.72 & $-$0.86 & 0.43\\
 2S & 50 & 2.55 & $-$1.20 & $-$0.60\\
 \enddata
 \tablenotetext{a}{From Ghosh \& Lamb 1992.}
\tablenotetext{b}{1G: Optically thick, gas-pressure
dominated (GPD) disk; 1R: Optically thick,
radiation-pressure dominated (RPD) disk; 2B:  
Two-temperature, optically thin GPD disk with
Comptonized bremsstrahlung; 2S: Two-temperature,
optically thin GPD disk with Comptonized soft photons
(for references to these disk models see Ghosh \& Lamb
1992).} \end{deluxetable} }

{\small 
 \begin{deluxetable}{cccc}
 \tablewidth{340pt}
 \tablecaption{Best-Fit Parameters for the
Magnetospheric Beat-Frequency Model}
 \tablehead{Source & $\nu_{s}$ (Hz)\tablenotemark{a}
& Normalization ($A_1$) & Exponent ($\alpha/\lambda$)}
 \startdata
 GX~17$+$2  & 293.5$\pm$5.5 & 365$\pm$1 &
  0.21$\pm$0.01\\
 GX~5$-$1   & 312.5$\pm$9.2 & 378$\pm$3 &
  0.16$\pm$0.02\\
 GX~340$+$0 & 324.5$\pm$9.3 & 389$\pm$4 &
  0.15$\pm$0.03\\
 Cyg~X-2    & 324.4$\pm$44.9 & 393$\pm$1 &
  0.17$\pm$0.02\\
 Sco~X-1    & 304.7$\pm$1.3 & 354$\pm$1 &
  0.06$\pm$0.03 
 \enddata
 \tablenotetext{a}{Inferred from the frequency
separation of the kilohertz QPOs (see text).}
 \end{deluxetable}
}

{\small 
 \begin{deluxetable}{ccc}
 \tablewidth{350pt}
 \tablecaption{Maximum $I/M$ Ratios for Illustrative
Neutron Star Equations of State\tablenotemark{a}}
 \tablehead{E.O.S.\tablenotemark{b} & Mass
($M_\odot$) & $I_{\rm 45}/M$ }
 \startdata
 A  & 1.66 & 0.63 \\
AU  & 2.13 & 0.93 \\
FPS & 1.80 & 0.76 \\
 L  & 2.70 & 1.73 \\
 M  & 1.80 & 0.91  
 \enddata
 \tablenotetext{a}{The tabulated values are the
maximum ratios found in a computational survey of all
stable masses and spin rates for the equation of state
indicated, using the code developed by Cook, Shapiro,
\& Teukolsky\markcite{CST94} (1994).}
 \tablenotetext{b}{A, AU, and FPS are realistic
equations of state; M illustrates the effects of a
strongly repulsive tensor interaction; L is an
extremely stiff equation of state produced by a simple
relativistic mean-field theory. For a recent
discussion of these equations of state, see Miller,
Lamb, \& Cook (1998).}
 \end{deluxetable}
 }


\begin{references}

\reference{ASP89} Angelini, L., Stella, L., \& Parmar,
A.\  N.\ 1989, ApJ, 346, 906

\reference{Aetal82} Alpar, M.\,A., Cheng, A.\,F., 
Ruderman, M.\,A., \& Shaham, J.\ 1982, Nature, 300, 728

\reference{AS85} Alpar, M.\,A., \& Shaham, J.\ 1985,
\nat, 316, 239

\reference{AY97} Alpar, M.\,A., \& Yilmaz, A.\ 1997, 
New Astr., 2, 225

\reference{CST94} Cook, G.\ B., Shapiro, S.\ L., \& 
Teukolsky, S.\ 1994, ApJ, 424, 823

\reference{CZC98} Cui, W., Zhang, S.\ N., \& Chen, 
W.\ 1998, \apj, 492, L53

\reference{Fetal97} Ford, E.\,C., Kaaret, P., Tavani,
M., Barret, D., Bloser, P., Grindlay, J., Harmon,
B.\,A., Paciesas, W.\,S., \& Zhang, S.\,N.\ 1997, ApJ,
475, L123

\reference{FK98} Ford, E.\,C., \& van der Klis, M.\
1998, \apj, 506, L39

\reference{FWH96} Finger, M.\ H., Wilson, R.\ B., \& 
Harmon, B.\ A.\ 1996, ApJ, 459, 288

\reference{G96} Ghosh, P.\ 1996, \apj, 459, 244

\reference{GL79} Ghosh, P.\ \& Lamb, F.\ K.\ 1979, 
\apj, 232, 259

\reference{GL92} ---------.\ 1992 in X-ray Binaries
and Recycled Pulsars, ed. E.P.J. van den Heuvel and
S.A. Rappaport (Dordrecht: Kluwer), 487

\reference{HK89} Hasinger, G., \& van der Klis, M.\
1989, \aap, 225, 79

\reference{Hetal98} Homan, J., van der Klis, M., 
Wijnands, R., Vaughan, B., Kuulkers, E.\ 1998, \apj, 
499, L41

\reference{I96} Ipser, J.\ R.\ 1996, ApJ, 458, 508

\reference{Jetal98} Jonker, P., Wijnands, R., van der
Klis, M., Psaltis, D., Kuulkers, E., \& Lamb, F.\ K.\
1998, ApJ, 499, L191

\reference{K95} Kuulkers, E.\ 1995, Ph.\,D.\ Thesis, 
University of Amsterdam

\reference{KK96} Kuulkers, E., \& van der Klis, M.\ 
1996, A\&A, 315, 567

\reference{KK98} ---------.\ 1998, A\&A, 332, 845

\reference{Ketal94} Kuulkers, E., van der Klis, M.,
Oosterbroek, T., Asai, K., Dotani, T., van Paradijs,
J., \& Lewin, W.\,H.\,G.\ 1994, A\&A 1994, 289, 795

\reference{Ketal96} Kuulkers, E., van der Klis, M.,
Oosterbroek, T., van Paradijs, J., \& Lewin, 
W.\,H.\,G.\ 1997, MNRAS, 287, 495

\reference{KKV96} Kuulkers, E., van der Klis, M., \&
Vaughan, B.A. 1996, A\&A, 311, 197

\reference{88} Lamb, F.\ K.\ 1988, in Physics of
Compact Objects, ed. N.\ E.\ White \& L. G. Filipov
(Adv. Space Res., 8), p.~421

\reference{89} Lamb, F.\ K.\ 1989, in Proc. 23 ESLAB
Symp. on X-ray Astronomy, ed. N.\ E.\ White (ESA
SP-296), p.~215

\reference{Letal85} Lamb, F.\ K., Shibazaki, N., Alpar,
M.\ A., \& Shaham, J.\ 1985, \nat, 317, 681

\reference{ML98} Markovi\'c, D., \& Lamb, F.\ K.\
1998,  ApJ, 507, 316

\reference{Metal98} M\'endez, M., van der Klis, M., van
Paradijs, J., Lewin, W.\,H.\,G., Vaughan, B.\,A.,
Kuulkers, E., Zhang, W., Lamb, F.\,K., \& Psaltis, D.\
1998, \apj, 494, L65

\reference{M99} Miller, M.\,C.\ 1999, \apj, in press 
(astro-ph/9809235)

\reference{MLC98} Miller, M.\,C., Lamb, F.\ K., \& 
Cook, G.\ 1998, \apj, 509, 793

\reference{MLP98} Miller, M.\,C., Lamb, F.\ K., \& 
Psaltis, D.\ 1998, \apj, 508, 791

\reference{MS99} Morsink, S.\,M., \& Stella, L.\ 1999,
\apj, in press (astro-ph/9808227)

\reference{PBK99} Psaltis, D., Belloni, T., \& van 
der Klis, M.\ 1999, \apj, in press

\reference{PL98} Psaltis, D., \& Lamb, F.\ K.\ 1998, 
in Neutron Stars and Pulsars, ed.\ N.\ Shibazaki, N.\
Kawai, S.\ Shibata, \& T. Kifune (Tokyo: Universal 
Academy Press), 179

\reference{PLM95} Psaltis, D., Lamb, F.\ K., \&
Miller,  G.\ S.\ 1995, ApJ, 454, L137

\reference{PLZ96} Psaltis, D., Lamb, F.\ K., \&
Zylstra, G.\ J.\ 1996, Proceedings of the NATO ASI:
Solar and Astrophysical MHD Flows, Astr.\ Lett.\ \&
Comm., 34, 377

\reference{Petal98} Psaltis, D., M\'endez, M., 
Wijnands, R., Homan, J., Jonker, P., van der Klis, 
M., Lamb, F.\,K., Kuulkers, E., van Paradijs, J., \& 
Lewin, W.\,H.\,G.\ 1998, ApJ, 501, L95

\reference{RS82} Radhakrishnan, V., \& Shrinivasan, 
G.\ 1982, Curr.\ Sci., 51, 1096

\reference{SL89} Shibazaki, N.\ \& Lamb, F.\ K.\ 1987, 
\apj, 318, 767

\reference{SMB97} Smith, B.\ A., Morgan, E.\ H., \& 
Bradt, H.\ 1997, ApJ, 479, L137

\reference{S97} Stella, L.\ 1997, talk presented at
the  HEAD meeting of the AAS, Estes Park, November 4--7

\reference{SV98} Stella, L., \& Vietri, M.\ 1997, in
The Active X-ray Sky, eds.\ L.\ Scarsi, H.\ Bradt, P.\
Giommi, \& F.\ Fiore, Nuclear Physics B Proc.\ Ser.,
135

\reference{SV98} ---------.\ 1998, \apj, 492, L59

\reference{SZS97} Strohmayer, T.\ E., Zhang, W.,
Swank,  J.\ H.\ 1997, ApJL, 487, L77

\reference{Setal96} Strohmayer, T.\ E., Zhang, W.,
Swank,  J.\ H., Smale, A., Titarchuk, L., Day, C., \&
Lee, U.\  1996, ApJL, 469, L9

\reference{Setal98} Strohmayer, T.\ E., Zhang, W.,
Swank, J.\ H., White, N.\,E., \& Lapidus, I.\ 1998,
\apj,  498, L135

\reference{vdK89} van der Klis, M.\ 1989, \araa, 27,
517

\reference{vdk98} ---------.\ 1998, in The Many Faces
of Neutron Stars, eds.\ R.\ Buccheri, J.\ van
Paradijs, \& M.\,A.\ Alpar (Dordrecht: Kluwer), 337
(astro-ph/9710016)

\reference{vdK85} van der Klis, M., Jansen, F., van
Paradijs, J., van den Heuvel, E.\ P.\ J., \& Lewin, 
W.\ H.\ G.\ 1985, \nat, 316, 225

\reference{Ketal97} van der Klis, M., Swank, J.\ H.,
Zhang, W., Jahoda, K., Morgan, E.\,H., Lewin, W.\ H.\
G., Vaughan, B., \& van Paradijs, J.\ 1996, ApJ, 469,
L1

\reference{Ketal97} van der Klis, M., Wijnands, R.,
Horne,  K., \& Chen, W.\ 1997, ApJ, 481, L97

\reference{VS98} Vietri, M., \& Stella, L.\ 1998, 
ApJ, 503, 350

\reference{WZ97} White, N.\ E., \& Zhang W.\ 1997, 
ApJ, 490, L87

\reference{Wetal97} Wijnands, R., Homan, J., van der
Klis, M., M\'endez, M., Kuulkers, E., van Paradijs,
J., Lewin, W.\ H.\ G., Lamb, F.\ K., Psaltis, D.,
Vaughan, B.\ 1997, ApJ, 490, L157

\reference{Wetal98a} Wijnands, R., Homan, J., van der 
Klis, M.,  Kuulkers, E., van Paradijs, J., Lewin, W.\
H.\ G., Lamb, F.\ K., Psaltis, D., Vaughan, B.\ 1998a, 
ApJ, 493, L87

\reference{Wetal98b} Wijnands, R., M\'endez, M., van 
der Klis, M., Psaltis, D., Kuulkers, E., \& Lamb, 
F.\,K.\ 1998b, \apj, 504, L35

\reference{WK97} Wijnands, R., \& van der Klis, M.\
1997,  ApJ, 482, L65

\reference{WK98} ---------.\ 1998, in Accretion
Processes in Astrophysical Systems, Proceedings of the
8th Annual Astrophysics Conference in Maryland, eds.
S.\ S.\ Holt \& T.\ Kallman, 381

\reference{Wetal96} Wijnands, R., van der Klis, M.,
Psaltis, D., Lamb, F.\ K., Kuulkers, E., Dieters, S.,
van Paradijs, J., \& Lewin, W.\ H.\ G.\ 1996, \apj,
469, L5

\reference{WKP98} Wijnands, R., van der Klis, M., \& 
van Paradijs, J.\ 1998, in The Hot Universe, IAU
Symp.\  188, in press

\reference{Zetal97} Zhang, W., Lapidus, I., Swank, J.\ 
H., White, N.\ E., \& Titarchuk, L.\ 1997, IAU Circ.\ 
6541

\reference{ZSS98} Zhang, W., Strohmayer, T.\,E., \& 
Swank, J.\,H.\ 1998, \apj, 500, L167

\end{references}
\end{document}